\documentstyle[11pt,newpasp,twoside,epsf]{article}
\begin{document}
\title{Atomic Polarization and the Hanle Effect}
\author{Javier Trujillo Bueno\footnote{1}}
\affil{Instituto de Astrof\'\i sica de Canarias, 38200 La Laguna, 
Tenerife, Spain. (e-mail: jtb@iac.es)} 

\altaffiltext{1}{Consejo Superior de Investigaciones Cient\'\i ficas, Spain.}

\begin{abstract}
This article presents an introduction to optical pumping,
atomic polarization and the Hanle effect in weakly magnetized stellar
atmospheres. Although the physical processes and the theoretical framework
described here are of interest for applications in a variety of astrophysical
contexts (e.g. scattering polarization in circumstellar envelopes and
polarization in astronomical masers), the article focuses mainly on
the quest for understanding the physical origin
of the linearly polarized solar limb spectrum. 
It considers also the development of
the Hanle effect as a reliable diagnostic tool
for making feasible new advances in solar photospheric and chromospheric magnetism. 
Particular emphasis is given to a rigorous modeling of polarization phenomena
as the essential link between theory and observations. Some of the most recent
advances in this field are presented after carefully explaining
how the various radiation pumping mechanisms lead to atomic polarization
in the absence and in the presence of weak magnetic fields. 
 
\end{abstract}

\section{Introduction}

Probably the first thing I should point out is that the
``second solar spectrum'' is nothing but the observational signature of the ``order''
that exists in the atomic system\footnote{The ``second solar spectrum'' 
is a term adopted by Stenflo and Keller (1997)
to refer to the linearly polarized solar limb spectrum
which can be observed with spectropolarimeters that allow the detection
of very low amplitude polarization signals (with $Q/I$
of the order of $10^{-3}$ or smaller). 
Weak polarization signals in spectral lines had previously been observed in prominences
outside the solar limb (e.g. Lyot 1934; Hyder 1965; see also the review by Leroy 1989)
and on the solar disc close to the limb
(e.g. Redman 1941; Br\"uckner 1963; Wiehr 1978; Stenflo {\it et al.} 1983),
but most of the structural richness of the
linearly polarized spectrum had remained inaccessible.
The observations of Stenflo and coworkers 
with the polarimeter ZIMPOL (see also the atlas of Gandorfer, 2000)
have been confirmed (and extended to the full Stokes vector) by
Dittmann {\it et al.} (2001), Mart\'\i nez Pillet {\it et al.} (2001)
and Trujillo Bueno {\it et al.} (2001) using the Canary Islands telescopes.}.
This ``atomic organization'' is
what we call {\it atomic polarization} (i.e. the existence of population imbalances
among the sublevels of any given degenerate atomic or molecular level and/or the presence of
quantum interferences or coherences between any given pair of sublevels, even among those
pertaining to different levels). But, what is forcing the ions, atoms
and molecules of the stellar atmospheric plasma to behave this way?
As we shall see below, the atomic polarization is
the result of a transfer process of ``order'' from the radiation
field to the atomic system. It is thus natural that the logical structure
of this review article is as follows: quantification of the ``order''
of the radiation field (Section 3), quantification of the ``order''
of the atomic system (Section 4) and transfer of ``order''
from the radiation field to the atomic system (Section 5).

The interesting point for solar surface magnetism is that weak magnetic
fields (from 1 milligauss to 100 gauss, approximately) modify the atomic polarization
via the Hanle effect. However, in order to develop the Hanle effect as a reliable diagnostic
tool of weak magnetic fields it is extremely important to fully understand 
the reference case of non-magnetic scattering polarization. With this motivation in mind  
Section 6 presents, for the first time, results of multilevel 
scattering polarization calculations
taking fully into account all the relevant pumping mechanisms
and the transfer of atomic polarization among all the levels involved.
After considering in some detail the {\it unmagnetized} reference case
a few examples in Section 7
will show how weak magnetic fields (from a few milligauss to a few gauss)
modify the atomic polarization of multilevel atomic systems. 
Finally, Section 8 gives our concluding remarks. 
Section 2 is mainly dedicated to introducing the subject
to those readers approaching the Hanle effect for the first time. 

\section{Introduction to the Hanle effect}

This section is of introductory nature. Similar and additional information
may be found in Hanle's (1924) paper
on ``Magnetic Effects on the Polarization of Resonance Fluorescence'',
in the classical monograph of Mitchell and Zemansky (1934),
in Landi Degl'Innocenti's (1992) contribution to the first IAC Winter School,
in the book edited by Moruzzi and Strumia (1991), 
and in the papers by Landolfi \& Landi Degl'Innocenti (1986) and by
Manso Sainz \& Trujillo Bueno (1999).
Other reviews and keynote articles related to this
topic (including the determination of magnetic fields in solar prominences)
can be found in Landi Degl'Innocenti (1990), Stenflo (1994),
Faurobert-Scholl (1996) and Trujillo Bueno (1999).
For a discussion concerning the possibilities of the Hanle effect for
the detection of magnetic fields in stellar winds see Ignace {\it et al.} (1997). 

\subsection{Hanle's ``Doktorarbeit''}

The story began in 1923 when Wilhelm Hanle of G\"ottingen University 
published the first correct interpretation
of a previously observed phenomenon related to the effect
of a weak magnetic field on the {\it linear} polarization
of the spectral-line radiation scattered by mercury vapor 
illuminated {\it anisotropically} (see Hanle 1923; 1924). 
With respect to the linear polarization corresponding 
to the zero magnetic field case, the observed influence of a 
weak magnetic field (of the order of 1 gauss) was
a {\it rotation} of the plane of linear polarization
(observed experimentally by Hanle himself) and a {\it depolarization}
(clearly pointed out  previously by Wood and Ellett in 1923).
This so-called Hanle effect played a fundamental role in the
development of quantum mechanics, since it led to the introduction and
clarification of the concept of {\it coherent superposition}
of pure states (see Bohr 1924; Hanle 1924, 1925; Heisenberg 1925).
The Hanle effect is directly related to the generation of coherent 
superposition of degenerate Zeeman sublevels of an atom (or molecule)
by a light beam\footnote{A {\it coherent superposition} of two or more sublevels
of a degenerate atomic level is a quantum mechanical state given by a
linear combination of pure states of the atomic Hamiltonian.}. 
As the Zeeman sublevels are split by the magnetic field,
the degeneracy is lifted and the coherence (and, in general, also the population imbalances
among the sublevels) are modified. As we shall illustrate below
in the context of the solar polarized spectrum, 
this gives rise to a characteristic magnetic-field dependence of the 
linear polarization of the scattered light that is finding increasing
application as a diagnostic tool for magnetic fields
in astrophysics (see {\it Astrophysical Spectropolarimetry\/}, edited
by Trujillo Bueno, Moreno Insertis and S\'anchez 2001). 

\begin{figure}
\label{hanle1}
\plotone{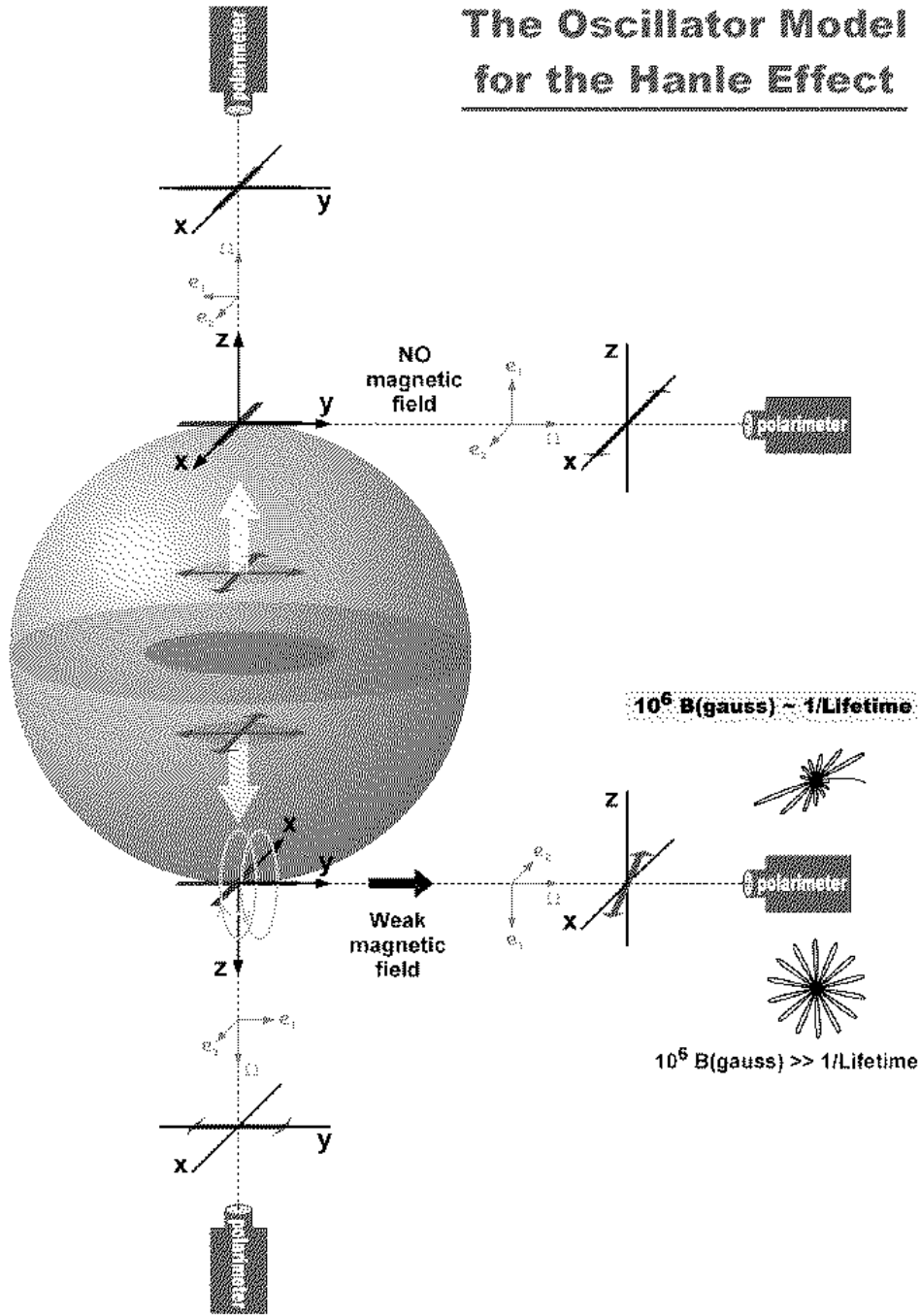}
\caption{The oscillator model for the Hanle effect.}
\end{figure}

\subsection{The oscillator model for the Hanle effect}

Figure 1 is aimed at introducing the most basic ideas behind scattering
polarization and the Hanle effect (see also Landi Degl'Innocenti, 1992). 
It is based on the classical interpretation
of the Hanle effect for a triplet-type transition originating between a ground level
with total angular momentum $J_l=0$ and an excited level with $J_u=1$. The atom is treated
as a negative charge oscillating with angular frequency $\omega_0$ 
and with a damping constant $\gamma$ given by the inverse of the 
lifetime of the excited atomic level
(e.g. the upper level of a resonance line transition has a lifetime
$t_{\rm life}{\approx}1/A_{ul}$,  $A_{ul}$ being
the Einstein coefficient for spontaneous emission). In solar-like atmospheres
$1/t_{\rm life} \ll {\Delta\omega}_{\rm line}$, with
${\Delta\omega}_{\rm line}$ the width of the spectral line under consideration.

As seen in Fig. 1, we are assuming that the atoms in the outer layers
of the solar atmosphere are being illuminated by an {\it unpolarized} radiation field,
which we are approximating by a unidirectional beam
propagating in the radial direction. The unpolarized character of this
radiation beam is indicated in Fig. 1 by two perpendicular {\it uncorrelated}
components of the electric field of the wave. Single-scattering events take place
and the light polarization is measured for both forward scattering and 
$90$ degree scattering as shown in Figure 1.

Consider first the observation of the ``north solar pole'', where we have assumed in Fig. 1
that there is {\it no} magnetic field (i.e. there is no Lorentz force influencing the
motion of the oscillating electron). Under these circumstances the atom can be
represented by three {\it independent} linear oscillators 
vibrating at angular frequency $\omega_0$ along the axes of the reference system.
As indicated in Fig. 1, only the $x$ and $y$ oscillators are excited by the incident
beam. The two excited oscillators radiate {\it independently} 
and decay radiatively with a damping constant $\gamma=1/t_{\rm life}$.
If we observe along the direction of the incident beam
(forward scattering case) we find that the radiation is obviously unpolarized. However,
observing at 90$^\circ$ one finds that the radiation is {\it linearly} polarized
along the $x$-axis, simply because the $y$-oscillator is seen pole-on. If we choose
the positive reference direction for the Stokes $Q$ parameter along the $x$-axis, we find
$Q=I$ and $U=0$. Note that the same conclusion is obviously reached if the vibration of
the $x$-oscillator is considered as that resulting from the {\it coherent
superposition} of two counter-rotating circular oscillators, which are
oscillating in phase with respect to each other at frequency $\omega_0$
in the $x$$-$$z$ plane. 

Consider now the observations of the ``south solar pole'', where we have
assumed in Fig. 1 the presence of a weak magnetic field {\it parallel} to the solar
surface and orientated along the $y$-axis. We have now to take into account
the influence of the Lorentz force on the motion of the bound electron.
The result is that the atom cannot be interpreted as three independent
linear oscillators, but as a linear oscillator parallel to the magnetic field
and two counter-rotating circular oscillators in the $x$$-$$z$ plane oscillating at 
angular frequencies $\omega_0+\omega_{\rm L}$ and $\omega_0-\omega_{\rm L}$, where
$\omega_{\rm L}\,=\,8.79\,\times10^6\,B({\rm gauss})$ is the Larmor angular frequency
(with $B({\rm gauss})$ indicating that the magnetic field is to be given in gauss).
The resulting trajectory of the electron in the $x$$-$$z$ plane is given by

\begin{equation}
x(t)=A\,e^{-{\gamma}t/2}{\rm cos}(\omega_{\rm L}t){\rm cos}(\omega_0t),
\end{equation}

\begin{equation}
z(t)=A\,e^{-{\gamma}t/2}{\rm sin}(\omega_{\rm L}t){\rm cos}(\omega_0t).
\end{equation}
The trajectory described by these equations is an oscillation at
frequency $\omega_0$, with an amplitude that decays exponentially
with a characteristic damping time given by $t_{\rm life}=1/\gamma$,
and such that its oscillation axis is rotating at frequency $\omega_{\rm L}$.
If the Zeeman splitting is sufficiently large so as to have
$\omega_{\rm L}\gg 1/t_{\rm life}$ 
(which can occur having still $\omega_{\rm L}\,{\ll}\,{\Delta\omega}_{\rm line}$)
the {\it oscillation axis} can rotate several times
before the oscillation amplitude is affected by the damping. The bound electron
describes in the $x$$-$$z$ plane the ``daffodil'' pattern shown in 
the lower part to the {r.h.s.} of Figure 1.
Under these circumstances we will see totally {\it unpolarized} radiation for the 90 degree
scattering case (i.e. for observation at the limb), 
but the maximum possible amount of linear polarization along the $y$-axis
for forward scattering (i.e. for disc-centre observation).

However, when the Zeeman splitting is similar to the natural width
of the atomic level (i.e. when $\omega_{\rm L}{\approx}1/t_{\rm life}$) the
oscillation axis rotates through an angle $\alpha$
within the characteristic damping time\footnote{The rotation angle 
$\alpha={\rm arctan}(U/Q)/2=\pm{\rm arctan}(2\Gamma)/2$. For an atomic level of
total angular momentum $J$ the parameter
$\Gamma=g_{J}\,\omega_{\rm L}\,t_{\rm life}$, with $g_{J}$ the Land\'e factor
(which is unity for the upper level with $J_u=1$
of the assumed triplet-type transition).}.
The bound electron describes in the $x$$-$$z$ plane the ``rosette'' shown in Figure 1.
For the 90 degrees scattering case (i.e. for observation at the limb) the $y$-oscillator
is seen pole-on and, therefore, the
observed polarization simply reflects the weighted average of the above-mentioned ``rosette'' pattern.
With respect to the previous {\it unmagnetized} ``north-pole case'', 
we now get (for observation at the limb)
a {\it depolarization} and a {\it rotation} of the polarization plane
(i.e. we now have a {\it smaller} Stokes $Q$ value and a {\it non-zero} Stokes 
$U$ signal).
This rotation is {\it counterclockwise} for a magnetic field 
pointing toward the observer, as in Fig. 1,
but {\it clockwise} if the magnetic field points
in the opposite direction. Accordingly, for a mixed-polarity magnetic field topology within
the spatio-temporal resolution element of the observations, the measured Stokes $U$ parameter would be
zero, while we would still be able to detect the {\it depolarization} effect by measuring Stokes $Q$.
For a disc-centre observation (forward scattering case) we would get 
(for the magnetic field orientation of Fig. 1) a net linear
polarization signal along the $y$-axis, but of smaller amplitude than that corresponding
to the previous $\omega_{\rm L}\gg 1/t_{\rm life}$ forward-scattering case.

Finally, it is also of interest to consider the ``south pole'' magnetized case, but
assuming that the weak magnetic field is now orientated along the $x$-axis. It is easy
to understand that all remains unchanged with regard to the forward
scattering case (with the exception that now the observed linear polarization is along
the $x$-axis, i.e. again along the magnetic field direction). However, for the limb observation
we would only see the depolarization effect (i.e. 
a Stokes $Q$ signal decreasing with the magnetic field, but $U=0$ always). 
Moreover, the Stokes $Q$ signal
does {\it not} vanish completely once the magnetic field becomes
sufficiently large so as to have $\omega_{\rm L} \gg 1/t_{\rm life}$. Figure 2
corresponds to the situation of a magnetic field 
which is always orientated along the $x$-axis. Note that a magnetic field parallel
to the solar surface and with an intensity in the saturated regime (i.e.
$\omega_{\rm L} \gg 1/t_{\rm life}$ or $\Gamma \gg 1$) leads to a
linear polarization amplitude for disc-centre observations that is
an order of magnitude {\it smaller} than that corresponding to the unmagnetized 
reference case ($\Gamma=0$) for an observation close to the solar limb ($\mu=0.1$).
Given the weakness of the ``predicted'' disc-centre polarization signals,
two spectral lines of possible interest for a
disc-centre Hanle-effect observational search of horizontal chromospheric
fields would be the Ca I 4227 \AA$\,$ line and the D$_2$ line of 
Ba II at 4554 \AA$\,$\footnote{For having in principle the possibility
of a positive detection the horizontal components of the 
chromospheric magnetic fields should {\it not} have 
a random azimuthal orientation within the spatio-temporal resolution element
of the observation.}.

\begin{figure}
\label{hanle2}
\plotone{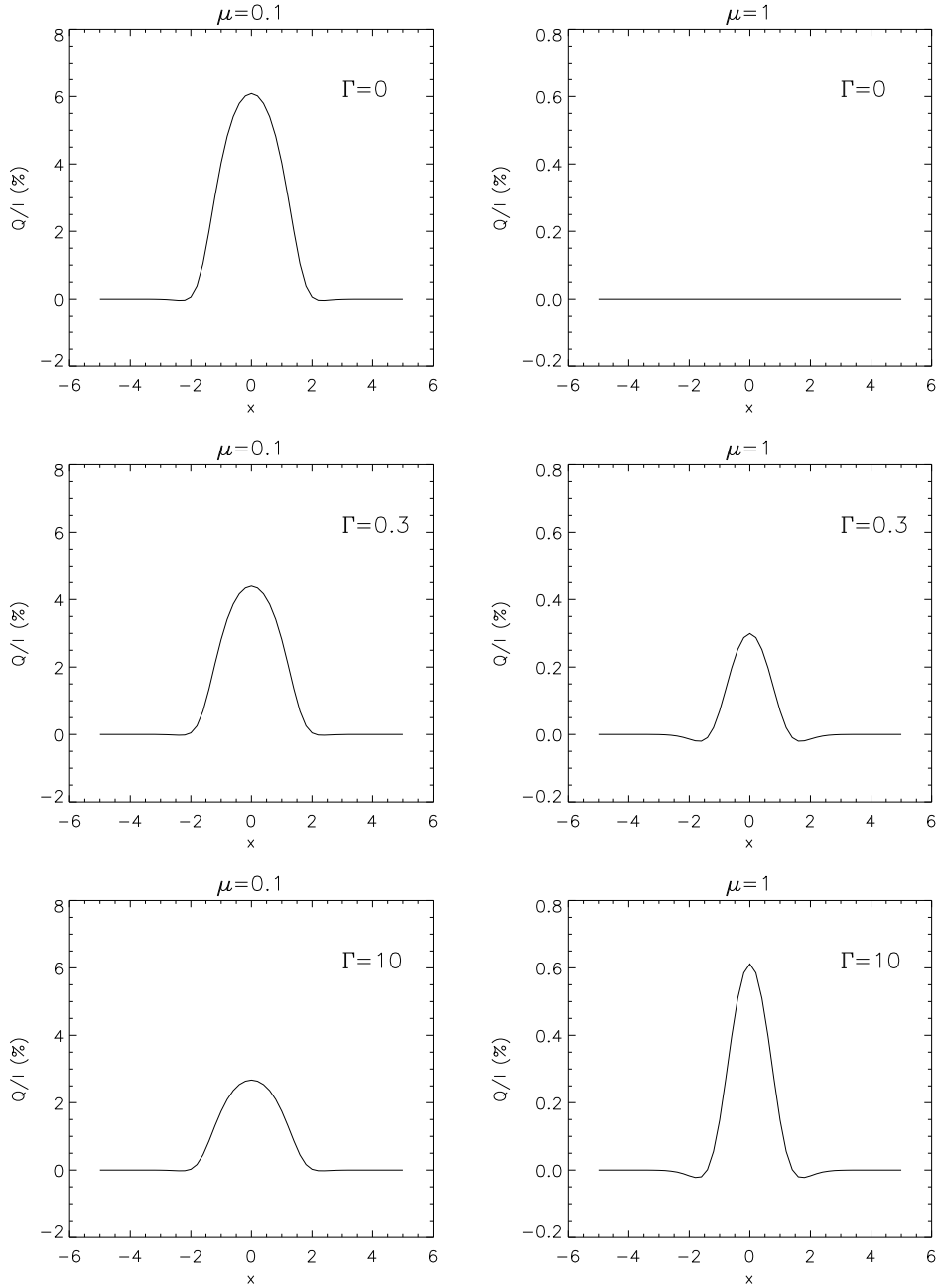}
\caption{Hanle-effect radiative transfer simulation for limb and
disc-centre observations. The figure shows the emergent fractional
linear polarization versus the line frequency in units of the Doppler width.
The positive reference-direction for the Stokes $Q$ parameter is
along the unit vector ${\bf e_2}$ of Figure 1.
Assumptions: triplet-type transition of a two-level
model atom with inelastic collisional destruction probability
$\epsilon=10^{-4}$ in an isothermal model atmosphere and neglecting depolarizing
elastic collisions. The magnetic field is parallel to the stellar surface
and orientated along the $x$-axis of Figure 1. The simulated observations are
as in Fig. 1 (for $\mu=1$, but for $\mu=0.1$ instead of $\mu=0$). The intensity 
of the magnetic field is quantified by 
$\Gamma=8.79{\times}10^6\,B{\rm (gauss)}\,g_J/A_{ul}$, with $g_J=1$.}
\end{figure}

\subsection{The basic Hanle-effect formula}

As we have seen, the Hanle effect produces a modification of the
{\it linear} polarization signals (quantified by the
Stokes $Q$ and $U$ parameters) with respect to the reference case
of non-magnetic scattering polarization. This occurs in a parameter domain
in which the {\it transverse} Zeeman effect is practically
ineffective. The Hanle effect is sensitive to magnetic fields
such that the corresponding Zeeman splitting is comparable to
the inverse lifetime of the lower or the upper atomic levels involved in the line transition
under consideration. The basic approximate formula to estimate
the {\it maximum} magnetic field intensity $B$ (measured in gauss)
to which the Hanle effect can be 
sensitive is\footnote{The natural formula to write is $\omega_{\rm L}\,g_J=
8.79{\times}10^6\,\,{\rm B}\,g_J\,{\approx}\,1/t_{\rm life}$. However,
by omitting the 8.79 factor we obtain an easier formula 
to remember (cf. Eq. 3), which informs us about the {\it maximum} magnetic
field intensity to which the Hanle effect can be sensitive.}

\begin{equation}\label{1}
10^6\,\,{B}\,\,g_{J}\,\,\approx\,\,1/t_{\rm life}\,\,,
\end{equation}
where $g_{J}$ and $t_{\rm life}$ are, respectively, the Land\'e factor 
and the lifetime of the atomic
level under consideration (which can be either the upper or the lower
level of the chosen spectral line transition). Depending on the
astrophysical plasma under consideration, on the intensity 
and orientation of its magnetic field and on the spectral line chosen,
we may find different Hanle-effect regimes. The most familiar one
to astrophysicists in general is the ``upper-level Hanle effect'',
in which only the upper-level coherences are modified by the action
of the magnetic field. However, we may also have
a ``lower-level Hanle effect'' regime in which only the lower-level coherences are 
sensitive to the field. The most general Hanle-effect takes place when the magnetic field affects
the coherences of both levels simultaneously. Obviously,
this dichotomy between an upper-level and a lower-level Hanle effect
is only a suitable one if there exists a sizeable difference
in the lifetimes of the lower and upper levels of the particular radiative transition. 
It certainly holds for solar spectral lines whose lower level
is either the {\it ground} or a {\it metastable} level, 
as happens, for example, with the D-lines of Na I
(Landi Degl'Innocenti 1998) or with the Mg I $b$ lines (Trujillo Bueno 1999).
It is important to note that in the solar atmosphere (whose $T_{\rm eff}{\approx}5800$ K) 
the lifetime of a ground or metastable level is typically 
two orders of magnitude {\it larger} than the lifetime of the upper level. 
Therefore, for typical solar spectral lines 
the upper-level Hanle effect would be sensitive to magnetic fields 
of between 1 and 100 gauss,
while the lower-level Hanle effect could in principle be used
for diagnosing much weaker fields, i.e. fields between $10^{-3}$ and 1 gauss.
Whether or not sub-gauss magnetic fields can actually exist in the
highly conductive solar plasma is a question that most solar plasma physicists
are inclined to answer negatively.  The main result I aim at demonstrating
in this article is that the ``enigmatic'' linear polarization signals 
observed by Stenflo {\it et al.} (2000) close to the solar limb 
in a number of chromospheric lines
are undoubtedly due to the presence of {\it atomic polarization}
in their {\it metastable-level} lower-levels. These observations
of scattering polarization on the Sun have been confirmed recently 
(and extended to the Stokes  $U$ and $V$ parameters) by Dittmann {\it et al.} (2001),
Mart\'\i nez Pillet {\it et al.} (2001) and by Trujillo Bueno {\it et al.} (2001) using 
different polarimeters attached to the Canary Islands solar telescopes.
How such  metastable-level atomic polarization can survive in the 
solar chromosphere is certainly a challenging question that I will also address in this paper. 
In particular, I will discuss briefly how a rigorous
theoretical interpretation of this type of spectropolarimetric 
observations is giving us decisive new clues about the topology and intensity of the magnetic
field of the ``quiet'' solar chromosphere.

\begin{figure}
\label{hanle3}
\plotone{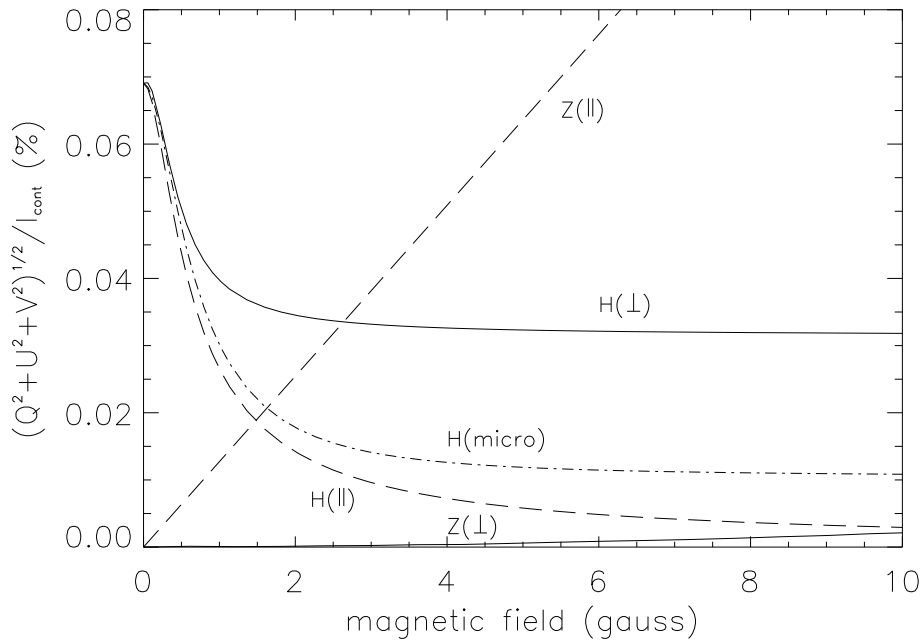}
\caption{The Hanle effect versus the Zeeman effect. The figure
shows the maximum polarization degree of a $J_l=0\,{\rightarrow}\,J_u=1$ resonance line 
(with $A_{ul}=10^7\ {\rm s}^{-1}$ and Doppler width ${\Delta{\nu}}_{\rm D}=10^{10}\ {\rm s}^{-1}$) 
versus the magnetic field intensity. The simulated observation is close to the limb ($\mu=0.1$)
of the model atmosphere of Figure 2. The polarization signals are due either to the Zeeman effect
(Z) or to the Hanle effect (H). The dashed--dotted line gives the sensitivity
of the Hanle effect to a {\it microturbulent and isotropic}
magnetic field. Note that for this kind of mixed-polarity scenario there is no Zeeman
signal. The two other cases are for a magnetic field {\it parallel}
to the stellar surface, and orientated either along the $x$-axis of Fig. 1 (curves
with the symbol $\perp$) or along the $y$-axis (curves with the symbol $\parallel$).
The polarimetric signals corresponding to the
transverse Zeeman effect (see the solid line labelled Z($\perp$)) have been
multiplied by a factor 10 to make them visible in the figure.
Note that for magnetic field intensities greater than 10 gauss (i.e. 
for the saturated Hanle-effect regime where $\Gamma{\gg}1$) the ``Hanle-effect signal''
is {\it only} sensitive to the {\it orientation} of the magnetic field vector, but
not to its intensity. This can occur for a Zeeman splitting which 
is still a very small fraction
of the width of the spectral line. 
(From Manso Sainz \& Trujillo Bueno 2001b).}
\end{figure}

\subsection{The utility of the Zeeman effect}

Before entering into details it is convenient to recall that
the Zeeman effect is most sensitive
in  {\it circular} polarization (quantified by the Stokes $V$ parameter),
with a magnitude that scales
with the ratio between the Zeeman splitting and the width
of the spectral line, and in a way such that the $V$ profile
changes its sign for opposite orientations of the magnetic field
vector. The longitudinal Zeeman effect in typical solar Fraunhofer
lines is of course rather insensitive to
sub-gauss magnetic fields. However, we should keep in mind that
today's state-of-the-art polarimeters are perfectly able to detect Stokes $V$ signals
corresponding to a flux density of only a 
few gauss (see, e.g., S\'anchez Almeida and Lites, 2000). Thus,
unless this flux density is well below 1 gauss in the spatio-temporal resolution element
of the observations, the longitudinal Zeeman effect itself
should be considered of complementary diagnostic value to the Hanle effect.
In any case, it is important to emphasize that the Hanle effect, contrary to the Zeeman effect,
does indeed work in any topologically complex weak-field scenario
(i.e. even if the net magnetic flux turns out to be exactly zero)
by producing a modification of the {\it linear} polarization signals
(with respect to the zero magnetic field reference case) that we can really ``measure''
if we succeed in rigorously modeling polarization signals
due to multiple-scattering processes in the presence of weak magnetic fields.
In conclusion, besides being useful for ``measuring'' magnetic fields in solar
prominences, the Hanle effect offers a promising diagnostic tool for
the investigation of weak {\it photospheric} fields with mixed polarities over small
spatial scales and for the exploration of chromospheric magnetic
fields (because chromospheric lines are generally broad and the magnetic fields
of the solar chromosphere relatively weak). 
Figure 3 illustrates what has just been pointed out
in relation to the diagnostic interest of the Hanle and Zeeman effects.
A suitable illustration of the fact that the Hanle effect only
operates in the Doppler core has been provided by Stenflo (1998)
for the case of a $J_l=0\,{\rightarrow}\,J_u=1$ line transition.

\section{How to quantify the order of the radiation field?}

In the standard multilevel radiative transfer problem (see Mihalas, 1978)
the only quantity related to the radiative line transitions that plays a role
in the statistical equilibrium equations is:

\begin{equation}
{\bar{J}^0_0}=\int {\rm d}x 
\oint \phi_x \frac{{\rm d}
\vec{\Omega}}{4\pi}\,{{I_{x \vec{\Omega}}}},
\end{equation}
where $x$ is the frequency measured from line centre in units of the Doppler width,
$\phi_x$ the absorption line shape,
$\vec{\Omega}$ the direction of propagation of the ray, 
and $I_{x \vec{\Omega}}$ the specific intensity
of the radiation field, i.e. the Stokes $I$ parameter.
Regardless of the dependence of the radiation field on
frequency and direction this
quantity (${\bar{J}^0_0}$) is always positive.
Note also that $B_{lu}{\bar{J}^0_0}$ 
gives the contribution 
of the absorption process (between a lower level $l$ and an upper
level $u$) to the natural width of the lower level ``$l$''
($B_{lu}$ is the Einstein coefficient for such an absorption process).
If this lower level is either the {\it ground} or a {\it metastable}
level, its lifetime 
is $t_{\rm life}\,{\approx}\,1/B_{lu}{\bar{J}^0_0}$
if and only if the line transition $l\,{\rightarrow}\,u$ is among the strongest ones.
This applies, for instance, to the lower-levels of the Mg $b$ lines,
to the lower levels of the Ca {\sc ii} IR-triplet, 
and to many other spectral lines in the solar spectrum.
All such lower levels are {\it metastable}. As a result, their lifetimes are about
two orders of magnitude larger than the lifetime of the upper levels
of the above-mentioned spectral lines.

However, in the most general polarization transfer case, 
there are eight additional radiation field quantities
that play a critical role in the statistical equilibrium equations
(see Landi Degl'Innocenti, 1983). They are the
spherical tensors ($\bar{J}^K_Q$) of the radiation field
(with $K=1,2$ and $-K\,{\le}\,Q\,{\le}\,K$) and they
are the quantities we choose to quantify the ``order''
of the radiation field:

\begin{equation}
{{\bar{J}^2_0}}=\int {\rm d}x 
\oint \phi_x \frac{{\rm d} \vec{\Omega}}{4\pi}
\frac{1}{2\sqrt{2}} \left[(3\mu^2-1){{I_{x \vec{\Omega}}}}+
3(\mu^2-1){Q_{x \vec{\Omega}}}\right]\,,
\end{equation}

\begin{equation}
{{\bar{J}^2_1}}=\int {\rm d}x 
\oint \phi_x \frac{{\rm d} \vec{\Omega}}{4\pi}
\frac{\sqrt{3}}{2} {{e^{i\chi}}}\sqrt{1-\mu^2}\,\left[ -\mu\left( {{I_{x \vec{\Omega}}}}+
{{Q_{x \vec{\Omega}}}}\right) -i{{U_{x \vec{\Omega}}}}\right]\,, 
\end{equation}

\begin{equation}
{{\bar{J}^2_2}}=\int {\rm d}x 
\oint \phi_x \frac{{\rm d} \vec{\Omega}}{4\pi}
\frac{\sqrt{3}}{2} {{e^{2i\chi}}}\,
\left[ \frac{1}{2}(1-\mu^2){{I_{x \vec{\Omega}}}}-
\frac{1}{2}(1+\mu^2){{Q_{x \vec{\Omega}}}}-i\mu {{U_{x \vec{\Omega}}}}\right]\,,
\end{equation}

\begin{equation}
{{\bar{J}^1_0}}=\int {\rm d}x \oint \phi_x \frac{{\rm d}
\vec{\Omega}}{4\pi}\,\sqrt{\frac{3}{2}}\mu\,{{V_{x \vec{\Omega}}}}\,,
\end{equation}

\begin{equation}
{{\bar{J}^1_1}}=-\int {\rm d}x \oint \phi_x \frac{{\rm d}
\vec{\Omega}}{4\pi}\,{{e^{i\chi}}}\,\sqrt{\frac{3}{2}}\sqrt{1-\mu^2}\,
{{V_{x \vec{\Omega}}}}\,,
\end{equation}
where the azimuthal angle $\chi$ and
$\mu={\rm cos}\theta$ specify the orientation
of each ray of direction $\vec{\Omega}$. The reference direction for the Stokes $Q$
and $U$ parameters is situated in the plane perpendicular to $\vec{\Omega}\,$ and
lies in the plane containing $\vec{\Omega}$ and the $z$-axis. The ${{\bar{J}^K_{-Q}}}$
components can be obtained easily from 
${{\bar{J}^K_{-Q}}}=(-1)^Q[{{\bar{J}^K_Q}}]^*$ (with $Q>0$, 
and where the symbol ``*'' means complex conjugation.).

In a weakly polarizing
medium like the ``quiet'' solar atmosphere (cf. S\'anchez Almeida and Trujillo Bueno 1999)
these radiation field tensors are essentially related with the degree
of ``order'' of Stokes $I$,
with the degree of ``order'' of Stokes $Q$ and $U$, and 
with the degree of ``order'' of Stokes $V$.
By degree of ``order'' I mean, essentially, degree of anisotropy,
degree of breaking of the axial symmetry,
and degree of ``lack of antisymmetry'' of the Stokes $V$ parameter.
An example of a completely ``disordered'' radiation field
is that of a black body. It is isotropic, unpolarized and it has
axial symmetry around any chosen direction in space. It is straightforward
to show that its only non-zero radiation field tensor is 
the mean radiation intensity: $\bar{J}^0_0(\nu)=B_{\nu}$. 
However, the radiation field of a stellar
atmosphere definitely has a degree of ``order'', which can be suitably quantified
by the above-mentioned radiation field tensors. 

\begin{figure}
\label{hanle4}
\plotone{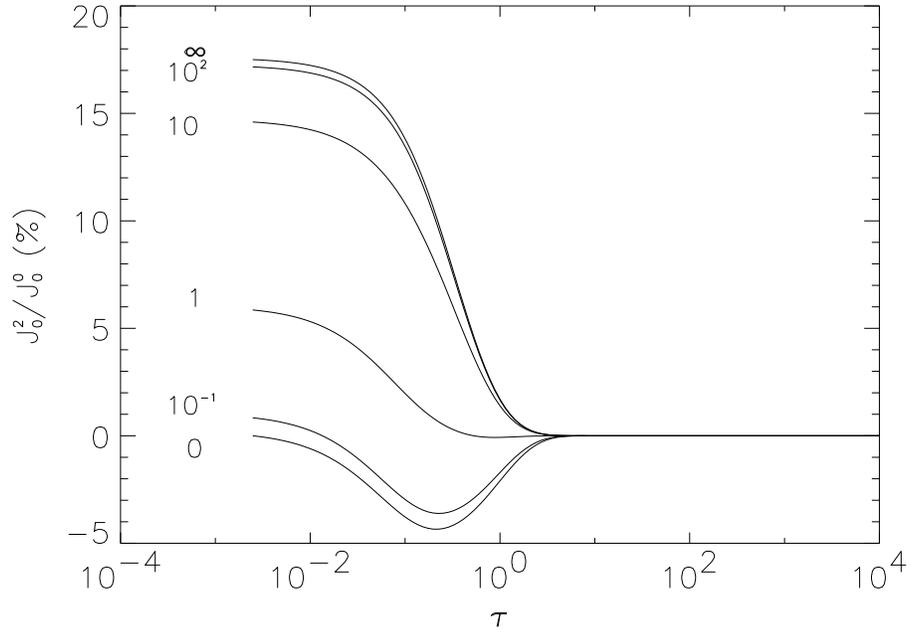}
\caption{The anisotropy factor in Milne--Eddington atmospheres.}
\end{figure}
\begin{figure}
\label{hanle5}
\plotone{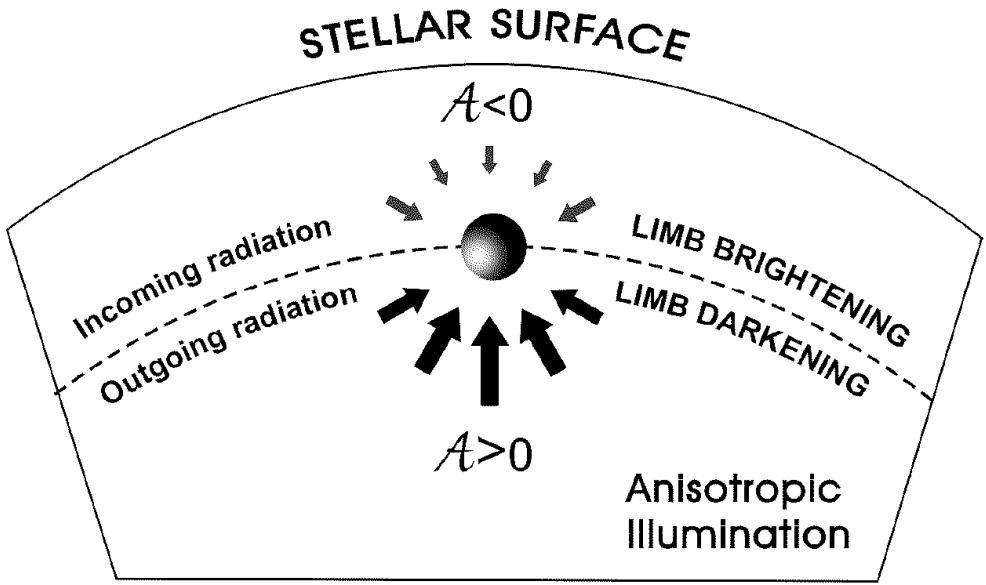}
\caption{The anisotropic illumination in a stellar atmosphere.}
\end{figure}

As an illustrative example, let us calculate
the degree of anisotropy in an unmagnetized Milne--Eddington atmosphere
(i.e. a medium in which the source function is $S\,=\,a\,+\,b\,\tau$,
with $\tau$ the optical depth). Figure 4 shows, for various values of $b/a$,
the variation with $\tau$ of the anisotropy factor 

\begin{equation}
{\cal{A}}\,=\,{{{\bar{J}}^2_0}\over{{\bar{J}}^0_0}},
\end{equation}
where $\bar{J}^0_0$ and $\bar{J}^2_0$ are given by Eqs. (4) and (5), respectively.
To obtain the results of Fig. 4 one has to solve the transfer equation in order to 
calculate the variation with $\mu={\rm cos}{\theta}$
of the Stokes $I$ parameter ($\theta$ is the angle between the ray direction
and the $z$-axis of the reference system, which we choose here along the 
normal to the stellar surface). Note that ${\cal{A}}\,{\le}\,0$
for the case of an atmosphere with no gradient in the source function 
(i.e. the curve with $b/a=0$ corresponding to $S=a$). 
The anisotropy factor (${\cal{A}}$)
is essentially negative in atmospheres with very small $b/a$ values
(see, for example, the curve with $b/a=0.1$), while ${\cal{A}}\,{\ge}\,0$
in atmospheres with sufficiently large $b/a$ ratios (see, for example, the curves
with $b/a {\ge} 1$). Note also that the larger $b/a$ the larger the
anisotropy factor. 

These results can be understood intuitively as follows. 
First, note that the tensor ${\bar{J}}^2_0$
(cf. Eq. 5) is dominated by the first term having the Stokes $I$ contribution. Therefore,
predominantly {\it vertical} rays (i.e. those with $\mu>1/\sqrt{3}$) make
{\it positive} contributions to $\bar{J}^2_0$, while predominantly
{\it horizontal} rays (i.e. those with $\mu<1/\sqrt{3}$) make {\it negative} 
contributions\footnote{The angle corresponding to $\mu=1/\sqrt{3}$
is known as Van Vleck's angle.}.
Second, as illustrated in Fig. 5, in a stellar atmosphere 
the {\it outgoing} intensities diminish with decreasing $\mu$ (i.e. they show
{\it limb darkening}), while the {\it incoming} intensities augment with decreasing $\mu$
(i.e. they show {\it limb brightening}). In other words, 
in a stellar atmosphere the outgoing radiation
is predominantly vertical, while the incoming radiation is predominantly
horizontal. Thus, the outgoing intensities tend to produce
{\it positive} contributions to $\bar{J}^2_0$, while the incoming intensities
tend to produce {\it negative} contributions. Therefore, there is  competition.
It wins the subset of intensities (outgoing or incoming) having the largest
variation with $\mu$. If $b/a=0$ the source function is constant ($S=a$)
and the outgoing intensities have no $\mu$ dependence. This implies
that ${\cal{A}} {\le} 0$ because the incoming intensities always show a dependence with
$\mu$, even for a constant-property atmosphere. This is simply due to the
presence of the stellar surface. However, as soon as $b/a$ becomes
``large enough'' the {\it limb darkening} of the outgoing intensities
becomes more important than the {\it limb brightening} of the incoming intensities
and the anisotropy factor ($\cal{A}$) becomes positive. 
It is straightforward to show that, for a given temperature
gradient of the model atmosphere, the corresponding source function gradient
is larger the bluer we go in the spectrum. This is the main reason
why the anisotropy factor generally increases as we
move in the spectrum from the IR toward the UV.
 
The anisotropy factor (cf. Eq. 10) is a fundamental quantity
in scattering polarization. It is also of interest to note that its
possible values are bounded as dictated by the following expression:

\begin{equation}
-{1\over{2\sqrt{2}}} \,\, {\le} \,\, {\cal A} \,\, {\le} \,\, {1\over{\sqrt{2}}}.
\end{equation}
In the chosen reference frame (with the $z$-axis along the radial stellar direction),
the largest $\cal{A}$ value corresponds to an
illumination coming from purely vertical (radial) radiation beams and the lowest
one to a purely horizontal radiation field.  

The remaining radiation field tensors are more subtle, but
their physical meaning can also be understood intuitively.
The tensors $\bar{J}^2_Q$ (with $Q=1$ and $Q=2$) are complex quantities.
They quantify the breaking of the axial
symmetry of the Stokes $I$, $Q$ and $U$
 parameters. If the physical properties of the
stellar atmosphere model depend only on the radial direction,
the ensuing $\bar{J}^2_Q$ tensors (with $Q=1$ and $Q=2$)
are zero unless the magnetic field
is {\it inclined} with respect to the radial direction. 
Figure 6 shows the variation with depth 
in a stellar model atmosphere of the real and imaginary parts of the
$\bar{J}^2_Q$ tensors normalized to $\bar{J}^0_0$. Note that
$\bar{J}^K_Q/\bar{J}^0_0 \ll 1$, i.e. that the degree of ``order''
of the radiation field which interacts with the atoms of the stellar
atmospheric model is weak. This must imply that the degree of ``order''
of the atomic system (i.e. the atomic polarization) has also to be  weak. 
These considerations can suitably be taken into account in order to 
facilitate some analytical insight into and the numerical solution of the general equations.

\begin{figure}
\label{jbar}
\plotone{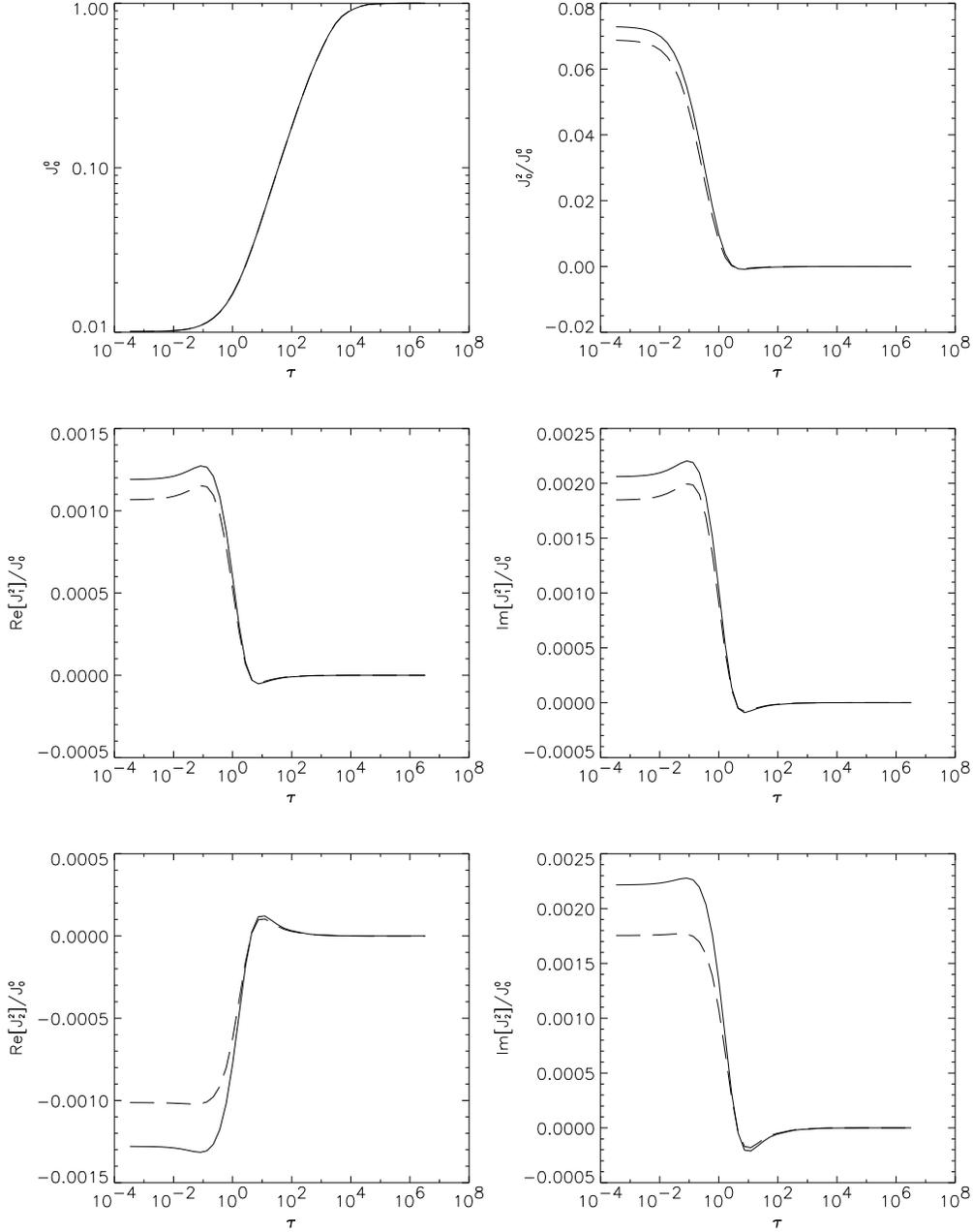}
\caption{The tensors $\bar{J}^2_Q/\bar{J}^0_0$ corresponding
to the self-consistent solution in the same
stellar atmospheric model of Figure 2. The assumed
horizontal magnetic field has $\Gamma=1$ and is
orientated at $30^\circ$ with respect to the $x$-axis of Figure 1.  
The solid lines give the exact
result. The dashed line of the panel for $\bar{J}^2_0/\bar{J}^0_0$
neglects the Stokes $Q$ contribution of Eq. (5). The dashed lines
of the remaining panels neglect 
the contribution of the Stokes $I$ parameter to Eqs. (6) and (7).
Therefore, the Stokes $I$ parameter is essential for calculating $\bar{J}^2_0$,
while the Stokes $Q$ and $U$ parameters cannot be neglected (in general) when
calculating $\bar{J}^2_1$ and $\bar{J}^2_2$.}
\end{figure}

The tensors $\bar{J}^1_Q$ are related to 
the circular polarization. Note that $\bar{J}^1_Q=0$ if the Stokes $V$
profile is perfectly antisymmetric, and that $\bar{J}^1_1=0$ if there
is no azimuthal $\chi$ dependence in the Stokes $V$ parameter. We thus need
to have net circular polarization in order to permit 
 non-zero values for the $\bar{J}^1_Q$ tensors.

The most general situation in which all the above-mentioned
radiation field tensors are non-zero is the case of a magnetized
stellar atmosphere with macroscopic velocity gradients. In this article
I will consider static stellar atmospheres for which the $\bar{J}^1_Q$
tensors can be assumed to be zero. As shown in Section 5, I will focus
mainly on the generation of atomic {\it alignment}, which is directly related
to the $\bar{J}^2_Q$ tensors and with the Hanle effect (i.e. with
the modification of the {\it linear polarization} signals due to weak magnetic fields).
In a future occasion I will address the issue of the generation of atomic
{\it orientation}, which is intimately related to the $\bar{J}^1_Q$ tensors
and with the modification of the {\it circular polarization} signals.

\section{How to quantify the ``order'' of the atomic system?}

The ``order'' of the atomic system
(i.e. the {\it atomic polarization}) 
can be suitably quantified via the
atomic density operator ($\rho^{\rm A}$)
of quantum mechanics. This operator
is represented in the basis {$\{\vert{n}\rangle \}$} 
of energy eigenvectors
via a matrix called the {\it atomic density matrix} whose elements are:

\begin{equation}
\rho^A_{ij}\,=\,
\langle {i}\vert{\rho^{\rm A}}\vert{j}\rangle .
\end{equation}
As clarified below, the diagonal elements 
(${{\rho}}^{A}_{ii}$) quantify the population 
of the state $\vert{i}\rangle$, while the non-diagonal
elements (${{\rho}}^{A}_{ij}$, with $i\neq j$)
quantify the degree of {\it quantum interference}
(or {\it coherence}) between the 
states $\vert{i}\rangle$ and $\vert{j}\rangle$.

Let us give some information regarding
the exact meaning of the atomic density matrix. 
First, note that the physical system under consideration is a volume element 
of a stellar atmosphere, which can be considered as being composed of 
three subsystems: the atoms (A) which emit, absorb, or scatter the 
radiation, the material perturbers (P) capable of influencing the 
excitation state of the atoms through inelastic and elastic collisions,
and the radiation field (R) itself. The quantum-mechanical Hamiltonian
operator ${\rm {\bf H}}$ of the whole system can be written, in the Schr\"odinger
representation, in the form ${\rm {\bf H}={\bf H}_0+{\bf V}}$, where
${\rm {\bf H}_0}={\rm {\bf H}_{\rm A}}+{\rm {\bf H}_{\rm R}}+{\rm {\bf H}_{\rm P}}$ 
is the unperturbed Hamiltonian made up of the
sum of the energies of the atoms, radiation and perturbers, while
${\rm {\bf V}}$  accounts for the various possible interactions among atoms,
perturbers and radiation. Second, recall that there are two
reasons which lead to the introduction of probabilities 
in the description of the state of a physical system such as that just defined.
One is the quantum-mechanical uncertainty related to the measurement process. The 
other is due (as in classical statistical mechanics) to our lack of 
complete information on the initial state of the system. This means 
that its description cannot be given in terms of a pure
state $\vert{\Psi}\rangle$, but through a statistical mixture of states (with the 
probability $p_{k}$ of finding the system in the state vector 
$\vert{\Psi}_{k}\rangle$). To incorporate into the quantum mechanical formalism the incomplete 
information we posses about the state of the system, the mixed state 
density operator (i.e. 
${{\rho}}=\sum_{k}{p}_{k}\vert{\Psi}_{k}\rangle\langle\Psi_{k}\vert$) was introduced (von 
Neumann, 1927; but see Fano, 1957). The essential point is that
the microscopic interactions
of our atomic subsystem with the perturbers and the radiation field
drive the atomic subsystem into a {\it mixed state} which can be described by 
the density operator but not by a pure state.

To grasp the usefulness of the atomic density operator, consider a 
complete set $\{\vert{n}\rangle \}$ of eigenvectors of the atomic Hamiltonian 
(i.e. ${\bf H}_{A}\vert{n}\rangle =E_{n}\vert{n}\rangle $, with $E_{_n}$ the energy 
eigenvalue corresponding to $\vert{n}\rangle$). To be able to make physical 
predictions about measurements bearing only on the atomic subsystem 
(A), the density operator ${{\rho}}^{A}$ of this subsystem is introduced
(i.e. ${{\rho}}^{A}$ = Tr$_{\rm P}$ (Tr$_{\rm R}{{\rho}})$, where the symbols 
Tr$_{\rm P}$ and Tr$_{\rm R}$ mean
the traces over the perturbers (P) and the radiation field (R) 
coordinates respectively; note that the trace of a matrix is the sum 
of its diagonal elements).
Calculating the matrix elements 
${{\rho}}_{ij}^{A}$ of ${{\rho}}^A$ in the set {$\{\vert{n}\rangle \}$} one finds:

\begin{itemize}

\item ${{\rho}}^{A}_{ii}$  
represents the average probability of finding the system in the state 
$\vert{i}\rangle $ (which implies that ${{\rho}}^{A}_{ii}$
quantifies the population of the state $\vert{i}\rangle $).

\item ${{\rho}}^{A}_{ij}$ (with $i\neq j$)
gives account of the interference effects between the 
states $\vert{i}\rangle$ and $\vert{j}\rangle$, which can appear simply because 
each state $\vert{\Psi}_{k}>$ of the statistical mixture is (in general) 
given by a coherent linear superposition of the states 
{$\{\vert{n}\rangle \}$}. As mentioned above, the 
non-diagonal elements of ${\rho}^{A}$ are called coherences.

\end{itemize}

The density operator ${{\rho}}$ of the total system (atoms, perturbers and radiation)
contains all physically significant
information on the system. In fact, the expectation
value of any observable, described in the Schr\"odinger
picture by a Hermitian operator $\bf O$, is given by (cf. Fano, 1957):

\begin{equation}
\langle{\bf O}\rangle(t)={\rm Tr}({\rho}{\bf O})={\rm Tr}({\rho}_{I}{\bf O}_{I}),
\end{equation}
where ${\bf O}_{I}$ and ${\rho}_{I}$ are the corresponding operators
in the interaction (or Heisenberg) picture 
of quantum mechanics (see any suitable advanced textbook).
As indicated by this expression, the expectation value of an observable
has the same structural form in the two quantum mechanical approaches
(i.e. in the Schr\"odinger and in the Heisenberg representations). However,
the Heisenberg picture has the advantage of forcing the time dependence of
wave functions to arise solely from the effect of the perturbing Hamiltonian ${\rm {\bf V}}$,
thus facilitating a perturbation approach.

Equation (13) is one of the two basic equations on which
the density-matrix polarization transfer theory is based.
The other fundamental equation is the equation of motion
for the density operator ${\rho}_{I}$, which 
reads\footnote{It is often called the Liouville equation in the interaction picture.
The corresponding equation in the Schr\"odinger picture is identical to Eq. (14),
but with ${\rho}$ instead of ${\rho}_{I}$ and
with the total Hamiltonian ${\rm {\bf H}}$ in place of ${\rm {\bf V}_{I}}$.
It can be easily derived from Schr\"odinger's equation.}

\begin{equation}
i{h\over{2\pi}}{{\partial{{\rho}_{I}}}\over{{\partial{t}}}}=[{\rm {\bf V}_{I}},{\rho}_{I}].
\end{equation}

The combination of Eqs. (13) and (14) leads to an exact equation for
$d\langle{\bf O}\rangle /dt$,
which can be used to derive the time evolution of physical quantities
(e.g. the time evolution of the diagonal and non-diagonal elements 
of the {\it atomic} density matrix ${{\rho}}^{A}$
due to the coupling of the atomic system A
with the radiation field R considered as a reservoir; cf. Cohen-Tannoudji, 1977).
As reviewed in more detail by Trujillo Bueno (1990), the above-mentioned 
exact equation for the time evolution of the observables
is the starting point in Landi Degl'Innocenti's
(1983) QED derivation of the statistical equilibrium and Stokes-vector transfer equations.

Finally, we should mention that there are two relevant
representations of the atomic density operator:
the {\it standard} (already introduced above)
and the {\it spherical statistical tensor}
representation (Omont, 1977; Blum, 1981). Let us introduce them
in terms of the basis of eigenvectors of the angular momentum
($|{\alpha}JM>$, with $\alpha$ indicating the quantum numbers of a given 
term---e.g. $\alpha=nLS$ if the atom is described by the $L-S$ coupling scheme).

In the {\it standard representation} the coherences
between magnetic sublevels pertaining to the same $J$-level
are given by\footnote{We may also have
coherences between sublevels of {\it different} $J$ levels.
In some cases (e.g. the H and K lines of Ca {\sc ii}) such coherences may have 
some observable effects (see Stenflo, 1980).}:

\begin{equation}
\rho^{\rm A}_{\alpha{J}}(M,M^{'})\,=\,
\langle {\alpha}JM\vert{\rho^{\rm A}}\vert{\alpha}JM^{'}\rangle ,
\end{equation}
while the total population of the $J$-level is
\begin{equation}
n_J=\sum_{M}\,{\rho^{\rm A}_{\alpha{J}}}(M,M),
\end{equation}
where $\rho^{\rm A}_{\alpha{J}}(M,M)$ is the
population of the sublevel with magnetic quantum number $M$.

In the {\it spherical statistical tensor}
representation the density-matrix elements are denoted
by the symbol $\rho^K_Q$ (with $K=0,...,2J$ and $-K {\le} Q {\le} K$).
The $\rho^K_Q$ elements are given by the following linear combinations
of the density-matrix elements of the standard representation (cf. Omont, 1977):

\begin{equation}
{\rho^K_Q}({\alpha}J)=\sum_{MM'}(-1)^{J-M}{\sqrt{2K+1}}
\left( \begin{array}{ccc}
J&J&K \\
M&-M^{'}&-Q
\end{array} \right)
{\rho^{\rm A}_{{\alpha}J}}(M,M^{'}),
\end{equation}
where the 3-j symbol is defined as indicated by any
suitable textbook on quantum mechanics.

For instance, for a level with total angular momentum $J=1$

\begin{equation}
\rho^0_0={1\over{\sqrt{3}}}
[\rho^{\rm A}_1(1,1)+\rho^{\rm A}_1(0,0)+\rho^{\rm A}_1(-1,-1)]={1\over{\sqrt{3}}}\,n_{J=1},
\end{equation}

\begin{equation}
\rho^1_0={1\over{\sqrt{2}}}[\rho^{\rm A}_1(1,1)-\rho^{\rm A}_1(-1,-1)],
\end{equation}

\begin{equation}
\rho^1_1=-{1\over{\sqrt{2}}}[\rho^{\rm A}_1(1,0)+\rho^{\rm A}_1(0,-1)],
\end{equation}

\begin{equation}
\rho^2_0={1\over{\sqrt{6}}}[\rho^{\rm A}_1(1,1)-2\rho^{\rm A}_1(0,0)+\rho^{\rm A}_1(-1,-1)],
\end{equation}

\begin{equation}
\rho^2_1=-{1\over{\sqrt{2}}}[\rho^{\rm A}_1(1,0)-\rho^{\rm A}_1(0,-1)],
\end{equation}

\begin{equation}
\rho^2_2=\rho^{\rm A}_1(1,-1).
\end{equation}

Note that the $\rho^K_Q$ elements with $Q=0$ are {\it real} numbers given
by linear combinations of the populations of the various Zeeman sublevels
corresponding to the level of total angular momentum $J$. The total
population of the atomic level
is quantified by ${\sqrt{2J+1}}{\rho^0_0}$, while the population imbalances
among such Zeeman sublevels are quantified by $\rho^1_0$ 
(i.e. by the $Q=0$ {\it orientation} coefficient)
and by $\rho^2_0$ (i.e. by the $Q=0$ {\it alignment} coefficient).
However, the $\rho^K_Q$ elements
with $Q {\ne} 0$ are {\it complex} numbers given by linear
combinations of the {\it coherences} between Zeeman sublevels whose magnetic quantum numbers
differ by $Q$. In fact, since the density operator is Hermitian, we have that
for each spherical statistical tensor component $\rho^K_Q$
with $Q>0$, there exists another component with $Q<0$ given by
$\rho^K_{-Q}=(-1)^Q[\rho^K_{Q}]^*$.
We thus have $(2J+1)^2$ density-matrix elements corresponding
to each level of total angular momentum $J$ (both in the standard and spherical
tensor representations).  

Both formalisms are totally equivalent. However, in the context
of scattering polarization and the Hanle effect, it is more convenient to work within
the framework of the spherical statistical tensor representation due to the
following reasons (cf. Landi Degl'Innocenti 1982).
The first advantage is directly related to the fact that the
density matrix elements on the basis of the eigenvectors of the
angular momentum depend on the reference system chosen to define
such eigenvectors. It turns out that the spherical statistical
tensors defined by Eq. (17) have simpler transformation laws
with respect to rotations of the coordinate system: their transformation
law involves just one rotation matrix instead of the product
of two rotation matrices. The second advantage is that the $\rho^K_Q$
elements themselves provide the most suitable way of quantifying,
at the atomic level, the information that we need to be able to calculate
all the ``sources'' and ``sinks'' of linear and circular polarization
within the medium under consideration (i.e. they have a clear physical
interpretation). For instance, if $\rho^2_0(J_u) {\ne} 0$ and $\rho^1_0(J_u) {\ne} 0$
we have local sources
of linear and circular polarization, respectively, even in the absence of magnetic fields.
In the Hanle effect regimes considered in this article 
the width of the spectral line is much larger than the Zeeman 
splitting and the Stokes $Q$ and $U$
components of the line emission vector are given by:

\begin{eqnarray}
  {\epsilon}_Q={\epsilon_{0}}\,w^{(2)}_{J_uJ_l}\Big{\{}\frac{3}{2\sqrt{2}}(\mu^2-1) {\rho}^2_0 -
  \sqrt{3}  \mu \sqrt{1-\mu^2} (\cos \chi
  {\rm Re}[{{\rho}}^2_1] - \sin 
  \chi {\rm Im}[{{\rho}}^2_1]) \nonumber 
\end{eqnarray}
\vspace{-0.2in}
\begin{eqnarray}
\hspace{1.9in} - \frac{\sqrt{3}}{2} (1+\mu^2) (\cos
  2\chi \, {\rm Re}[{{\rho}}^2_2]-\sin 2\chi \, {\rm Im}[{{\rho}}^2_2]) \Big{\}},
\end{eqnarray}
\vspace{-0.2in}
\begin{eqnarray}
\epsilon_U={\epsilon_{0}}\,w^{(2)}_{J_uJ_l}\sqrt{3} \,\Big{\{} \sqrt{1-\mu^2} ( \sin \chi
  {\rm Re}[{{\rho}}^2_1]+\cos \chi {\rm Im}[{{\rho}}^2_1]) \nonumber 
\end{eqnarray}
\vspace{-0.2in}
\begin{eqnarray}
\hspace{1.3in}  +
  \mu (\sin 2\chi \, {\rm Re}[{{\rho}}^2_2] + \cos 2\chi \, {\rm Im}[{{\rho}}^2_2]) \Big{\}},
\end{eqnarray}
where the $\rho^K_Q$
values are those of the {\it upper} level of the line transition under consideration,
$\epsilon_0=(h\nu/4\pi)A_{ul}{\phi_x}{\cal N}\sqrt{2J_u+1}$
(with $\cal N$ the total number of atoms per unit volume), $w^{(2)}_{J_uJ_l}$
is the symbol introduced by Landi Degl'Innocenti (1984) (which depends only
on $J_u$ and $J_l$), and where the orientation of the ray is specified by
$\mu={\rm cos}\theta$ (with $\theta$ the polar angle) and by the azimuthal
angle $\chi$. The elements $\eta_Q$ and $\eta_U$ of the absorption matrix
are given by
identical expressions (i.e. by $\eta_Q=\epsilon_Q$ and by $\eta_U=\epsilon_U$),
but with $\eta_0=(h\nu/4\pi)B_{lu}{\phi_x}{\cal N}\sqrt{2J_l+1}$ instead of $\epsilon_0$,
$w^{(2)}_{J_lJ_u}$ instead of $w^{(2)}_{J_uJ_l}$ and with the $\rho^K_Q$
values of the {\it lower} level of the line transition
(instead of those of the upper level).
Note that $\epsilon_Q$ and $\eta_Q$
depend on both the population imbalances ($\rho^2_0$)
and on the coherences ($\rho^2_Q$, with $Q=1,2$), while $\epsilon_U$ and $\eta_U$
depend {\it only} on the coherences. Recall also that the emission vector and the absorption
matrix are the two fundamental quantities which appear in the
Stokes-vector transfer equation.

Finally, an additional reason for the suitability of the spherical tensor representation
is that the limiting case in which polarization
phenomena are neglected (cf. Mihalas, 1978) can be obtained
simply by retaining only the terms with $K=Q=0$.

\section{Transfer of ``order'' from the radiation field to the atomic system}

There are two mechanisms which
can lead to transfer of ``order'' from the
radiation field to the atomic system: {\it upper-level} selective
population pumping and {\it lower-level} selective depopulation pumping.
The requirements are that the pumping light 
must be necessarily {\it anisotropic}, and/or {\it polarized}
and/or to have {\it spectral structure} over a frequency
interval $\Delta\nu$ smaller than the frequency
separation between the Zeeman sublevels. 

{\bf Upper-level population pumping} occurs when some {\it upper-state}
sublevels have more chances of being populated than others.
For instance, if an unpolarized light beam propagating
along the direction chosen as the quantization axis illuminates
a gas of two level atoms with $J_l=0$ and $J_u=1$, only
the transitions corresponding to ${\Delta}M={\pm}1$ are effective, so that
no transitions occur to the $M=0$ sublevel of the upper level. Thus,
in the absence of any relaxation mechanisms, the upper-level sublevels
with $M=1$ and $M=-1$ would be more populated than the $M=0$ sublevel and the 
fractional upper-level alignment coefficient 
$\sigma^2_0(u)={\rho}^2_0(u)/{\rho}^0_0(u)=1/{\sqrt{2}}$.

{\bf Lower-level depopulation pumping} occurs when some {\it lower-state} sublevels
absorb light more strongly than others. As a result an excess population
tends to build up in the weakly absorbing sublevels.  
For instance, if an unpolarized light beam propagating
along the direction chosen as the quantization axis illuminates
a gas of two level atoms with $J_l=1$ and $J_u=0$, only
the transitions corresponding to ${\Delta}M={\pm}1$ are effective, so that
no transitions can occur out of the $M=0$ sublevel of the lower level.
On the other hand, the spontaneous de-excitation from the upper level
populates with equal probability the three sublevels ($M=-1,0,+1$)
of the {\it lower} level. In the absence of any relaxation mechanisms,
the final result of this optical-pumping 
cycle is that all atoms will eventually 
be pumped into the $M=0$ sublevel of the {\it lower} level, and the medium will
become transparent (Happer, 1972).
Under such ideal laboratory conditions 
(illumination by a unidirectional light beam and absence of depolarizing mechanisms)
the fractional lower-level alignment coefficient 
is $\sigma^2_0(l)={\rho}^2_0(l)/{\rho}^0_0(l)=-{\sqrt{2}}$, i.e.
{\it a factor 2 larger} (in absolute value) {\it than
the fractional upper-level alignment corresponding to the previous triplet-line case}.

The Nobel laureate Alfred Kastler (1950)
was actually the first scientist to propose that optical pumping
under laboratory conditions can be used as a method
to change the relative populations of Zeeman sublevels
and of hyperfine levels of the ground state of atoms.
What we have been emphasizing over the last few years is that
the very same mechanism (lower-level depopulation pumping)
is operating in the atmospheres of the stars,
and that it constitutes {\it an essential physical
ingredient for understanding the second solar spectrum}. 

Lower-level depopulation pumping in solar-like atmospheres
was investigated in detail 
by Trujillo Bueno and Landi Degl'Innocenti (1997)
including depolarizing collisions and radiative transfer effects. 
We chose that particular type of line transition 
(i.e. $1\,\rightarrow\,0\,\rightarrow\,1$) to
point out clearly that lower-level depopulation pumping 
due to the anisotropy of the solar radiation field
can lead to sizeable amounts of ground-level atomic polarization and to emergent
linear polarization signals with amplitudes in the observable range. 
Soon afterwards, Landi Degl'Innocenti (1998) succeeded in deriving analytical expressions
for the Stokes-$Q$ component of the emission vector and of the absorption matrix
corresponding to the D$_1$ and D$_2$ lines of Na I, 
and could show by adjusting free parameters that a certain amount of atomic polarization in the
hyperfine components of the ground level of sodium leads to a remarkably good
fit of the complex fractional linear polarization 
pattern observed by Stenflo and Keller (1997), including the ``enigmatic'' line-centre peaks.  
A subsequent theoretical
investigation (Trujillo Bueno 1999) was aimed at demonstrating 
that {\it depopulation pumping}
in the solar atmosphere (and the induced lower-level atomic polarization)
can actually produce similar amounts of linear polarization
for some groups of line transitions (having different $J_l$ and $J_u$ values)
for which the simplified resonance line polarization
theory (which neglects the influence of {\it lower-level} depopulation pumping) 
predicts drastically different emergent polarizations. Based on this
result, I suggested
a possible explanation of the ``enigmatic'' linear polarization
amplitudes observed by Stenflo {\it et al.} (1983, 2000) in the Mg I $b$ lines
(see Trujillo Bueno 1999; Sections 6 and 7).
That theoretical investigation 
demonstrated that the presence of a sizeable amount of atomic polarization
in the {\it metastable} lower-levels of the Mg $b_1$ and $b_2$ lines would
explain in a natural way the 
similar polarization amplitudes observed
in the Mg $b$ lines (and also those observed in
the three lines of multiplet No. 3 of Ca I, which is also 
a $^3{\rm P}^0\,-\,^3{\rm S}$ multiplet). This result 
for the Mg $b$ lines was particularly encouraging
because, contrary to the case of the sodium D-lines, it is 
totally impossible to argue
that there might perhaps exist an alternative explanation of the observed polarization
peaks based on a multilevel scenario characterized by the absence
of lower-level polarization (see Trujillo Bueno 1999; Sections 5 and 6).

In addition to the two previous pumping mechanisms,
which allow the direct transfer of ``order''  from the radiation field to the atomic system,
there is an additional pumping process which also comes into play:
{\it repopulation pumping}. This pumping occurs either when the {\it lower-level}
is repopulated as a result of spontaneous decay of a {\it polarized} upper level
or when the {\it upper-level}
is repopulated as a result of absorptions of a {\it polarized} lower level.
As can be intuitively expected, the repopulation pumping rates are proportional
to the atomic polarization of the atomic levels.

\subsection{The two-level atom in the absence of magnetic fields}

The best thing we can do
to facilitate  understanding of all these pumping
mechanisms is to show them in action for a particularly illuminating example:
the scattering line polarization problem in a one-dimensional 
{\it unmagnetized} stellar atmosphere formed by
a gas of two level atoms with $J_l=J_u=1$ (cf. Trujillo Bueno 1999). For this case, the only
non-zero radiation field tensors are $\bar{J}^0_0$ and $\bar{J}^2_0$ and the unknowns
of the problem at each spatial point are simply 
$\rho^0_0(l)$, $\rho^2_0(l)$, $\rho^0_0(u)$ and $\rho^2_0(u)$.
The rate equations which govern the temporal evolution of the
alignment coefficients ($\rho^2_0$) of the upper and lower levels
can be deduced applying the density matrix polarization transfer theory
of Landi Degl'Innocenti (1983). We obtain:

\begin{eqnarray}
{{d}\over{d\,t}}\,{{\rho}^2_0(u)}\,=\,
-\left[ {B_{lu}\over2}{{{\bar{J}}^{2}_{0}}}
{{{\rho}^{0}_{0}}(l)}+
{{B_{lu}}\over{{2}}}{{{\bar{J}}^{0}_{0}}}
{{{\rho}^{2}_{0}}(l)}+
{B_{lu}\over{\sqrt{2}}}{{{\bar{J}}^{2}_{0}}}
{{{\rho}^{2}_{0}}(l)}\right] -A_{ul}\,{{{\rho}^{2}_{0}}(u)} \\
+C^{(2)}_{lu}{{{\rho}^{2}_{0}}(l)} 
-(C_{ul}+D_{u})
{{{\rho}^{2}_{0}}(u)},
\nonumber
\end{eqnarray}  

 \begin{eqnarray}   
{{d}\over{d\,t}}\,{{\rho}^2_0(l)}\,=\,
-A_{ul}{{{\rho}^{2}_{0}}(u)}
+\left[ {B_{lu}}{{{\bar{J}}^{2}_{0}}}
{{{\rho}^{0}_{0}}(l)}-
2{{B_{lu}}}{{{\bar{J}}^{0}_{0}}}
{{{\rho}^{2}_{0}}(l)}-
{B_{lu}\over{\sqrt{2}}}{{{\bar{J}}^{2}_{0}}}
{{{\rho}^{2}_{0}}(l)}\right] \\
+2C_{ul}
{{{\rho}^{2}_{0}}(u)}-2(C_{lu}+D_{l})
{{{\rho}^{2}_{0}}(l)}.
\nonumber
\end{eqnarray}   

These equations have contributions coming from {\it transfer} and
{\it relaxation} rates due to radiative and collisional terms.
I have neglected the stimulated emission terms. Assuming statistical
equilibrium (i.e. ${\rm d}{\rho^K_Q}/{\rm d}t=0$),
neglecting also inelastic collisions (i.e. the $C_{lu}$ and $C_{ul}$ terms)
and the terms containing the products ${{{\bar{J}}^{2}_{0}}}{{{\rho}^{2}_{0}}}$
(because in a weakly polarizing medium like the solar atmosphere 
${\bar{J}}^{2}_{0}/{\bar{J}}^{0}_{0}\ll 1$ 
and ${\rho}^{2}_{0}/{\rho}^{0}_{0}\ll 1$) we find:

\begin{equation}
{{{\rho^2_0}({\rm up})}}\,{\approx}\,
{{-1}\over{2\,(1+{{\delta}_{u}})}}\,\left[
{{B_{lu}{\bar{J}^2_0}}
\over{A_{ul}}}\,{{{\rho^0_0}({\rm low})}}
+{{B_{lu}{\bar{J}^0_0}}\over{A_{ul}}}\,{{{\rho^2_0}({\rm low})}}\right],
\end{equation}

\begin{equation}
{{{\rho^2_0}({\rm low})}}\,{\approx}\,
\,{1\over{2\,(1+{{\delta}_{l}})}}\,\left[
{{B_{lu}{\bar{J}^2_0}}
\over{B_{lu}{\bar{J}^0_0}}}\,{{{\rho^0_0}({\rm low})}}
-{{A_{ul}}\over{B_{lu}{\bar{J}^0_0}}}\,{{{\rho^2_0}({\rm up})}}\right],
\end{equation}
where ${\delta}_{u}=D_u/A_{ul}$ and ${\delta}_{l}=D_l/{B_{lu}{\bar{J}^0_0}}$,
with $D_u$ and $D_l$ the depolarizing rates due to elastic collisions.

The first term within the brackets of Eq. (28) is due
to the {\it upper-level} population pumping mechanism, while the first term
within the brackets of Eq. (29) is due to the {\it lower-level} depopulation
pumping mechanism. Both are given by the rate $B_{lu}{\bar{J}^2_0}{\rho^0_0}({\rm low})$;
the only difference is that the {\it upper-level} pumping contribution is 
multiplied by the lifetime of the upper level (i.e. by $1/A_{ul}$), while
the {\it lower-level} pumping contribution is multiplied
by the lifetime of the ground level (i.e. by $1/B_{lu}{\bar{J}^0_0}$, which
for optical line transitions in solar-like atmospheres is about two orders
of magnitude {\it larger} than the upper-level lifetime).

The second term within the brackets of Eq. (29) is due to the
{\it repopulation pumping} resulting from the spontaneous decay of the {\it polarized}
upper level, while the second term within the brackets of Eq. (28) is due to the
{\it repopulation pumping} resulting from the absorption process from the {\it polarized}
lower level. 

The previous two equations have been useful in identifying
the contributions corresponding to the various pumping mechanisms.
However, the relevant quantities of interest
to understanding the observed polarization amplitudes
are the fractional atomic polarization 
($\sigma^2_0=\rho^2_0/\rho^0_0$) of each level.
In fact, the emergent fractional linear polarization close to the solar limb
is approximately given by:

\begin{equation}
{Q\over{I}}\,\approx\,{\cal W}\,\sigma^2_0({\rm up})\,-\,
{\cal Z}\,\sigma^2_0({\rm low}),
\end{equation}
where ${\cal W}$ and ${\cal Z}$ are simply numbers given by the total angular momentum
values of the lower and upper atomic levels of the line 
transition under consideration\footnote{Note that ${\cal W}=w^{(2)}_{J_uJ_l}$ and
 ${\cal Z}=w^{(2)}_{J_lJ_u}$, where the symbol
$w^{(2)}_{JJ^{'}}$ is the one introduced by Landi Degl'Innocenti (1984).
For instance, for a $1{\rightarrow}1$ transition ${\cal W}={\cal Z}=-1/2$.}. 
In this formula
the $\sigma^2_0$ values have to be calculated at the spatial point (situated along the
line of sight) where the optical depth is unity. This formula (see Trujillo Bueno 1999)
may be considered as the generalization of the Eddington--Barbier relation to the 
non-magnetic scattering polarization case. (Note that I have rewritten it here in a way such
that the positive reference-direction for the Stokes $Q$ parameter is now chosen along the
line {\it perpendicular} to the radial direction through the observed point, and not
along the radial direction as it was chosen in the above-mentioned paper and in Eqs. 24 and 25.)

The equations for $\sigma^2_0({\rm up})$ and $\sigma^2_0({\rm low})$ can
be obtained easily after dividing Eqs. (28) and (29) by $\rho^0_0(l)$ and taking
into account that $\rho^0_0(u)/\rho^0_0(l)\,\approx\,B_{lu}{\bar{J}^0_0}/A_{ul}$.
(Note that this is a good approximation
for a weakly anisotropic medium
in which the inelastic collisional rates have been assumed to be 
much smaller than the radiative rates.) The result reads:

\begin{equation}
{{\sigma^2_0({\rm up})}}\,{\approx}\,
{{-1}\over{2\,(1+{\delta_u})}}\,\left[
{{{\bar{J}^2_0}}\over{{\bar{J}^0_0}}}
+{{\sigma^2_0({\rm low})}}\right],
\end{equation}

\begin{equation}
{{\sigma^2_0({\rm low})}}\,{\approx}\,
{1\over{2\,(1+{\delta_l})}}\,\left[
{{{\bar{J}^2_0}}\over{{\bar{J}^0_0}}}
-{{\sigma^2_0({\rm up})}}\right].
\end{equation}

We thus see that there exists a fascinating closed loop connecting
the lower-level and upper-level polarizations. The atomic polarization of the
upper level produced by the anisotropic illumination of the atoms
is modified (typically enhanced!) because of the atomic polarization
of the lower level, and vice versa. The only way to frustrate this remarkable communication 
between the two levels is by forcing one of the two levels to be
{\it totally} unpolarized, which can {\it only} occur in practice if either $\delta_u$
or $\delta_l$ turns out to be very much larger than unity. In fact, if we
assume that the atomic polarization of the lower level is totally destroyed
by elastic collisions (i.e. $\delta_l\,\rightarrow\,\infty$) and that 
the upper level is insensitive to such collisions (i.e. $\delta_u=0$)
we find ${\sigma^2_0({\rm up})}\,\approx\,-0.5\,({\bar{J}^2_0}/{\bar{J}^0_0})$,
in agreement with the self-consistent numerical results given by the dashed lines
of Fig. 1 of Trujillo Bueno (1999). However, if we assume that both atomic levels
are insensitive to elastic collisions (i.e. that $\delta_l=\delta_u=0$), we then
find ${\sigma^2_0({\rm up})}\,\approx\,-{\sigma^2_0({\rm low})}\,
\approx\,-\,{\bar{J}^2_0}/{\bar{J}^0_0}$,
in agreement with the self-consistent 
numerical result given in Fig. 3 of Trujillo Bueno (1999).

\subsection{The two-level atom in the presence of weak magnetic fields}

What do we obtain for the atomic polarization of the ground
and excited levels of a $1{\rightarrow}1$ transition
if we include the effect of a weak magnetic field? Obviously,
the situation becomes more complicated. First, the number
of {\it unknowns} at each spatial point is 12 instead of just
four. The density matrix elements whose values are sought are\footnote
{Actually, there are six additional unknowns which are related
with the atomic orientation (i.e. with the ${\rho^1_Q}$'s),
but these density-matrix elements are zero 
if the radiation field tensors ${\bar{J}^1_Q}=0$.}:  
$\rho^0_0(l)$, $\rho^2_0(l)$, the real and imaginary
parts of $\rho^2_1(l)$ and of $\rho^2_2(l)$,
$\rho^0_0(u)$, $\rho^2_0(u)$ and
the real and imaginary
parts of $\rho^2_1(u)$ and of $\rho^2_2(u)$.
Second, there are two ``natural'' choices for the direction
of the $z$-axis of the reference system
used to formulate the rate equations: ($a$) the magnetic field
reference frame in which the $z$-axis is aligned with the local magnetic field direction
and ($b$) the stellar radial direction itself.
For developing a general multilevel scattering polarization and Hanle-effect code
we prefer option $b$ (which is in fact the choice in Sections 6 and 7).
However, as done in the remaining part of this section,
it is convenient to formulate the rate equations in the magnetic field
reference frame if the interest lies in achieving some analytical insight.

The two following expressions are a particular case of the
equations derived by Landi Degl'Innocenti (1985).
They are valid for a $1{\rightarrow}1$ transition of a two-level atom
in a weakly anisotropic and magnetized stellar atmosphere in which
the effect of collisions is assumed to be negligible:

\begin{equation}
{\sigma^K_Q}({u})={\rho^K_Q}(u)/{\rho^0_0}(u)=-{ { {1\over{2}} (1+i{\Gamma_l}Q)\,+\,{1\over{4}}}\over
{(1+i{\Gamma_u}Q)(1+i{\Gamma_l}Q)\,-\,{1\over{4}}} } \,\,
(-1)^Q \, { {\bar{J}^K_{-Q}}\over{\bar{J}^0_0} },
\end{equation}

\begin{equation}
{\sigma^K_Q}({l})={\rho^K_Q}(l)/{\rho^0_0}(l)=(-1)^{K}{ { {1\over{2}} 
(1+i{\Gamma_u}Q)\,+\,{1\over{4}}}\over
{(1+i{\Gamma_u}Q)(1+i{\Gamma_l}Q)\,-\, {1\over{4}} } } \,\,
(-1)^Q \, { {\bar{J}^K_{-Q}}\over{\bar{J}^0_0} }.
\end{equation}

These formulae are valid for $K=1$ and $K=2$. 
They show that, once the fractional radiation
field tensors (i.e. ${ {\bar{J}^K_{-Q}}/{\bar{J}^0_0} }$)
are known {\it in the magnetic field reference frame},
the fractional atomic polarization of each level depend only on
the dimensionless quantities $\Gamma_u$ and $\Gamma_l$,
which are given by:

\begin{equation}
\Gamma_u=8.79\times10^6{\rm B}g_{J_u}/A_{ul},
\end{equation}

\begin{equation}
\Gamma_l=8.79\times10^6{\rm B}g_{J_l}/B_{lu}\bar{J}^0_0.
\end{equation}

The previous equations for ${\sigma^K_Q}({u})$ 
and ${\sigma^K_Q}({l})$ show that
{\it in the magnetic field reference frame}
the population imbalances (i.e. ${\sigma^K_0}$)
are {\it insensitive} to the magnetic field,
while the coherences (i.e. ${\sigma^K_Q}$ with $Q{\ne}0$)
are reduced and dephased as the magnetic field is increased.
Finally, note that magnetic fields such that
$\Gamma_l\,\approx\,1$ are very efficient in
depolarizing the ground level, which has a significant feedback on
the upper-level coherences, while fields
such that $\Gamma_u\,\approx\,1$ are very efficient in
depolarizing the excited level.

Some polarization diagrams illustrating the lower-level
and upper-level Hanle effects for single scattering
events in two-level atomic models have been presented by
Landolfi \& Landi Degl'Innocenti (1986). Similar type
of Hanle-effect diagrams, but taking into account radiative transfer effects
in isothermal stellar atmospheres, 
will be published by Manso Sainz \& Trujillo Bueno (2001b). 
 
\section{Multilevel scattering polarization}

The theoretical interpretation of scattering
polarization signals requires a 
calculation of the the atomic polarization considering complex
atomic and molecular systems. This is a very involved
{\it non-linear} and {\it non-local} problem which has been
solved with sufficient generality only recently (see below).
It consists in calculating, at each spatial grid-point
of the assumed astrophysical plasma, the density-matrix
elements which are {\it consistent} with the properties of the 
polarized radiation field generated within the medium.

From multilevel radiative transfer simulations
without polarization physics we know that the two-level atom
model is, with few exceptions, an unsuitable approximation for 
modeling the Fraunhofer spectrum. Obviously,
the same statement should be applied (but with much more emphasis) to 
the realm of the second solar spectrum, which is due to coherence transfer effects.
In order to exploit its rich diagnostic potential
(e.g. with the aim of exploring magnetoturbulence and/or to 
improve our understanding of chromospheric magnetism) it is crucial
to investigate the generation and transfer of atomic polarization 
considering realistic multilevel models and taking fully into account
all the above-mentioned pumping mechanisms in the presence (and absence) of the Hanle
effect induced by weak magnetic fields (e.g. from 1 milligauss to 100 gauss). During the 
last year Rafael Manso Sainz\footnote{He is now at the University of Florence, 
after having completed his Ph.D. thesis at the IAC.} 
and I have successfully developed a general multilevel
scattering polarization radiative transfer code, which is currently allowing us to
carry out this type of investigations via realistic numerical simulations. Our multilevel
scattering polarization code solves the following set of equations:

{\bf (a)} The rate equations giving the time evolution of the atomic density matrix
elements. In these equations the radiative rates are given in terms of the radiation field tensors
introduced in Section 3. For the moment we assume statistical equilibrium (SE)
(i.e. $d{\rho^K_Q}/dt=0$).

{\bf (b)} The Stokes-vector transfer equations where the components of the emission 
vector and those of the absorption matrix depend on the density-matrix elements themselves.
For the moment, we assume one-dimensional (1D) plane-parallel geometry, but we have also
developed suitable formal solution methods for considering spherical
circumstellar envelopes and realistic 3D model atmospheres with horizontal inhomogeneities. 

These SE and RT equations were derived from the principles of 
quantum electrodynamics by Landi Degl'Innocenti (1983).
This density-matrix polarization transfer theory is based on the Markovian assumption of
complete frequency redistribution (see a critical review of 
this CRD polarization transfer theory in Trujillo Bueno 1990).
For problems where the only significant coherences are those between
the sublevels of degenerate levels, this CRD theory should provide a physically
consistent description of scattering phenomena {\it if} the spectrum of the
pumping radiation is flat across a frequency range wider than the inverse lifetime
of the levels (Landi Degl'Innocenti {\it et al.} 1997). As we shall
see below, this seems to be a sufficiently good 
approximation for modeling the observed polarization in a number of spectral lines
of diagnostic interest (e.g. the Ca {\sc ii} 
IR triplet and the Mg $b$ lines)\footnote{In any case, it cannot be emphasized sufficiently
the importance of continuing the investigations to generalize
the density-matrix polarization-transfer theory to partial frequency
redistribution and to situations where the excitation of the atomic
system is sensitive to the spectral structure of the pumping radiation
(see Bommier 1997 a,b; Landi Degl'Innocenti {\it et al.} 1997).}.

Other scientists (e.g. Bommier and Sahal-Br\'echot 1978) had previously 
formulated the statistical equilibrium equations for the Hanle effect regime
applying also the theory of the master equation of a ``small system'' (the atom)
interacting with a ``big reservoir'' (the radiation field) (cf. Cohen-Tannoudji
1977; Cohen-Tannoudji {\it et al.} 1992). The
interest of Landi's theoretical work stems from the fact that he was able to achieve a general
{\it unified} theoretical derivation of both, the statistical equilibrium equations
{\it and} the radiative transfer equations\footnote{It is of historical 
interest to mention the
works of House (1970 a,b; 1971), who was among
the first to provide some significant contributions to the theory
of scattering polarization and the Hanle effect. See also the work of Litvak (1975) on polarization
of astronomical masers.}.
From the point of view of numerical
radiative transfer our coherency transfer code is based on the fast iterative
methods which the author of this article developed
for multilevel scattering polarization and Hanle effect applications 
(see Trujillo Bueno 1999; and note that the two-level atom scattering
polarization problem with lower-level depopulation pumping solved in that paper is 
also a non-linear and non-local problem of the same nature 
as the general multilevel case considered here).

\subsection{Three-level systems and the ``enigmatic'' Ca {\sc ii} 8662 \AA$\,$ line}

The quest to understand the physical origin of the second solar spectrum
will be facilitated if we first gain some physical insight on how 
the various pumping mechanisms
leads to atomic polarization in three-level atomic systems
under stellar atmospheric conditions. 

There are a number of schematic three-state
configurations which our laser physics colleagues like to include in their
textbooks: cascade, vee (${\bf {V}}$) and lambda (${\bf {\Lambda}}$).
The linkage pattern is $1\,\leftrightarrow\,2\leftrightarrow\,3$ for any ranking
of the relative energy values. Thus, the {\it cascade configuration} is the one in which
the energies increase with index $i$, $E_1\,<\,E_2\,<\,E_3$. The {\it vee configuration}
occurs when the central state (i.e. level 2) lies lowest in energy. A third possibility, the
{\it lambda configuration}, occurs when the central state lies highest in energy. 

Unfortunately, I do not have enough space here to present and discuss the results
of our coherency transfer calculations in stellar atmospheric models
for all these three-level systems. I have to make a choice: the {\it lambda configuration}
with $J_1=1/2$, $J_2=1/2$, and $J_3=3/2$.
My motivation behind this choice is two-fold. First, it is precisely the configuration
of the Raman scattering process, in which a photon of one frequency is absorbed
and a second frequency may be re-emitted. Second, this particular three-level
atomic system will help to understand physically our successful modeling of the
``enigmatic'' fractional linear polarization observed by Stenflo {\it et al.} (2000)
in the Ca {\sc ii} IR triplet (see Manso Sainz and Trujillo Bueno 2001a). To this end,
the atomic parameters of the $1\,\leftrightarrow\,2$ transition have been 
chosen to be identical
to that of the H line of Ca {\sc ii}, while those of the $2\,\leftrightarrow\,3$ transition
correspond to that of the 8662 \AA $\,$ line of the Ca {\sc ii} IR triplet. The observations
were considered ``enigmatic'' especially because this 8662 \AA $\,$ line should be,
according to Stenflo {\it et al.} (2000), intrinsically unpolarizable. However,
their observations show a polarized peak (see also Dittmann {\it et al.} 2001). 
They considered this line ``intrinsically unpolarizable'' because its upper level
(and also the lower level of the H-line) cannot harbour any atomic alignment
(because $J_2=1/2$ and calcium has no hyperfine structure).

\begin{figure}
\label{hanle6}
\plottwo{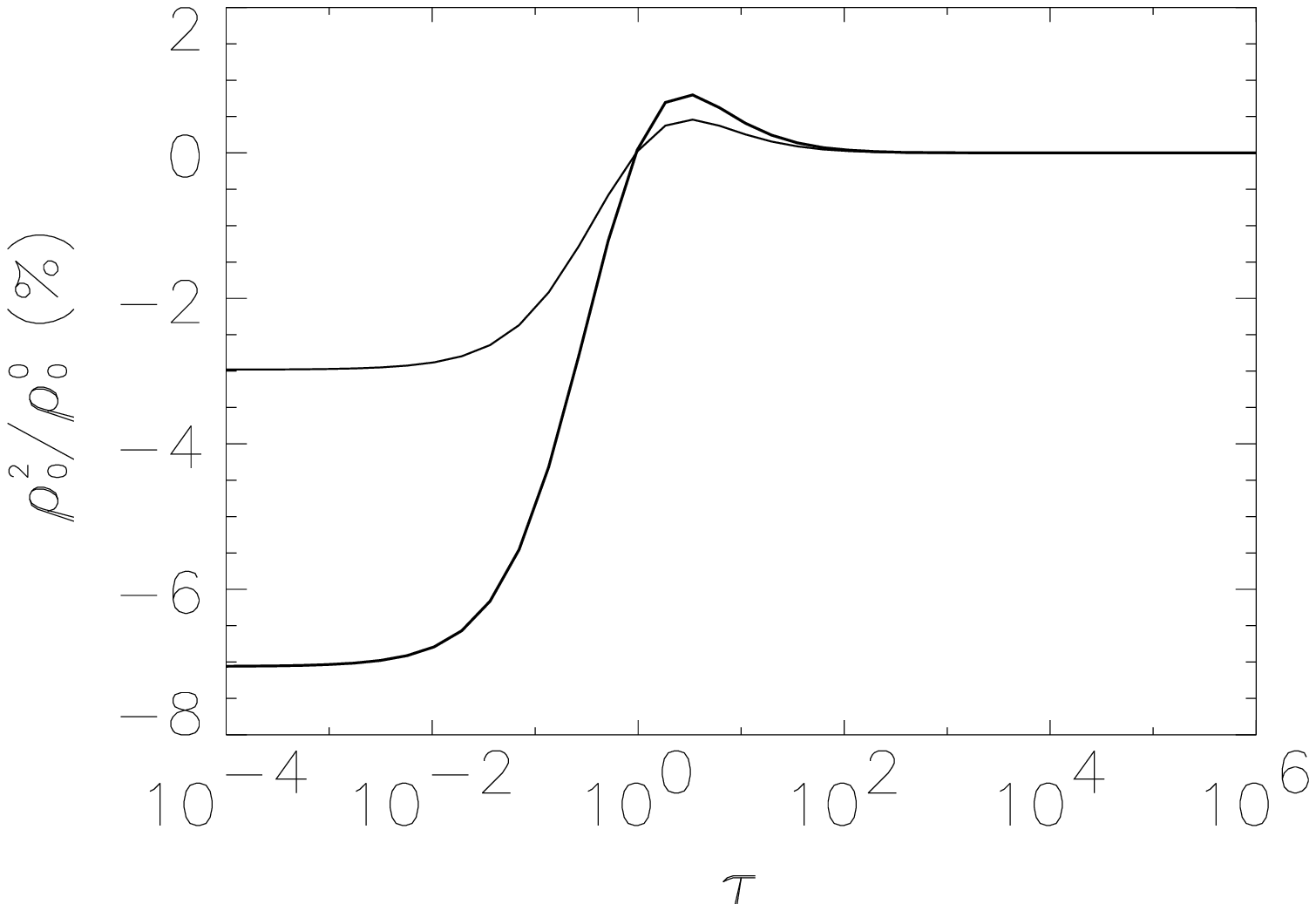}{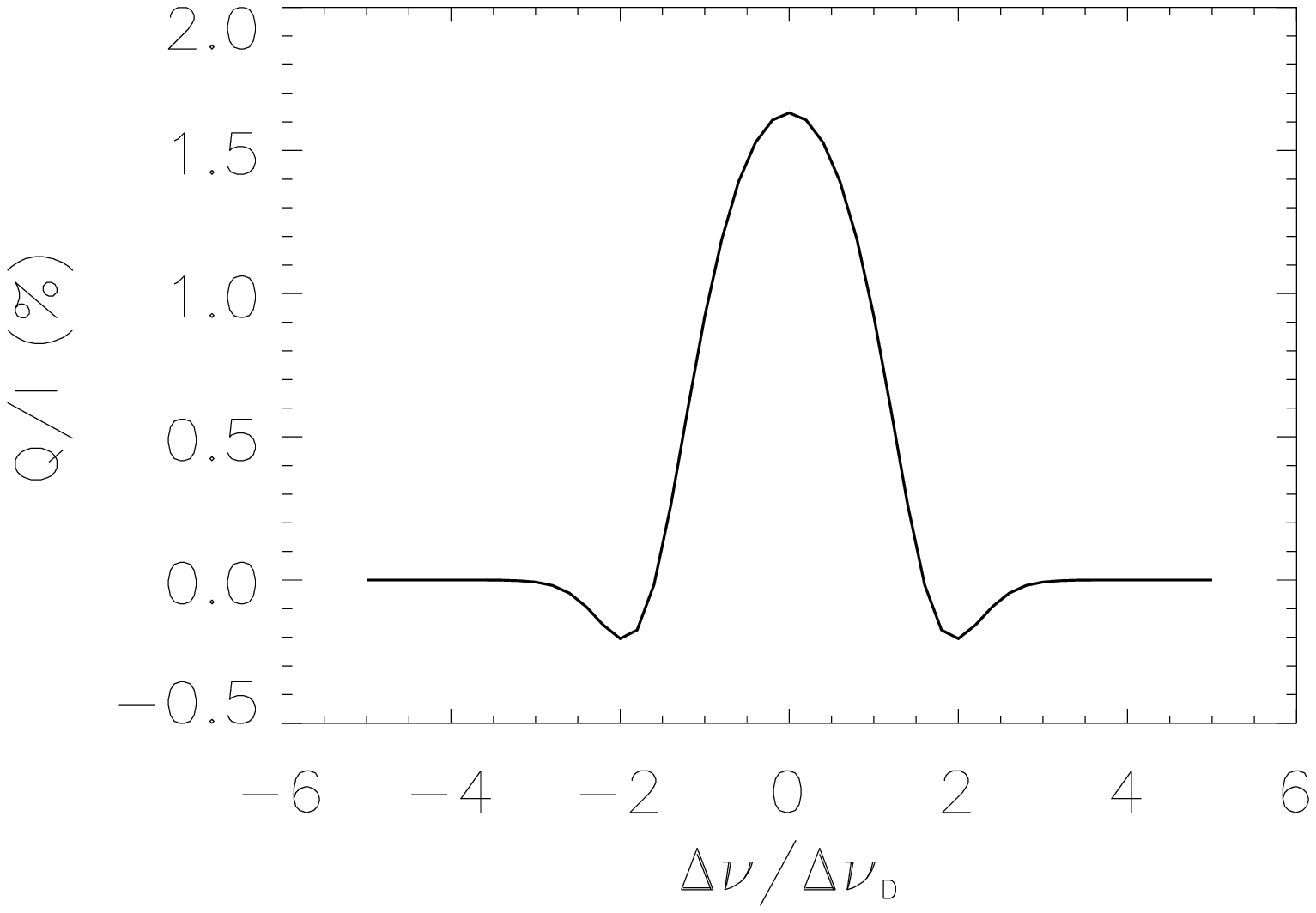}
\caption{Left: The fractional atomic polarization of the lower level of the 8662 \AA$\,$ line
versus its optical depth
(see the thin solid line yielding a surface value of $-3\%$). 
The thick solid line yielding a surface value of $-7\%$
shows $-{\bar{J}^2_0}/{\bar{J}^0_0}$.
Right: The corresponding emergent fractional linear polarization of 
the 8662 \AA$\,$ line at $\mu=0.1$.}
\end{figure}

Figure 7 shows the self-consistent results for the {\it unmagnetized} case. 
The thin line of the left panel gives the
variation with line optical depth of the fractional atomic polarization
of the lower level of the 8662 \AA $\,$ line (whose $J_3=3/2$). The thick line
shows the variation of the anisotropy factor at this wavelength (it actually
shows the variation of $-{\bar{J}^2_0}/{\bar{J}^0_0}$). We see that 
depopulation pumping due to the $2\,\leftrightarrow\,3$ transition 
at 8662 \AA $\,$ is able to generate sizeable unequal populations among their lower-level
sublevels. If one now uses our Eddington--Barbier relation
(i.e. Eq. 30) and takes into account that 
for a $J_l=3/2\,\rightarrow\,J_u=1/2$ transition ${\cal W}=0$ and ${\cal Z}{\approx}0.7$,
one ends up understanding the computed emergent fractional
linear polarization shown in the {\it rhs} figure. This emergent
polarization at $\mu=0.1$ is totally due to {\it dichroism} in the
{\it non-magnetized} solar model atmosphere (see Trujillo Bueno 1999; page 82).
In other words, the generation of the observed linear polarization in the 
Ca {\sc ii} 8662 \AA $\,$ line necessarily requires a transfer
process along the line of sight. This is needed in order to allow the Q-component of the
{\it absorption} matrix (i.e. $\eta_Q$, which is non-zero because the lower
level $^2{\rm D}_{3/2}$ is polarized) to make 
its influence on the emergent linear polarization \footnote{Note 
that the $Q$ component of the {\it emission}
vector is {\it zero} because the upper level cannot harbour any atomic alignment.}.
This is precisely the physical origin of the linear polarization
of the chromospheric Ca {\sc ii} 8662 \AA $\,$ line observed close to the ``quiet'' solar limb.

It is important to point out that the calculated line-core fractional polarization
amplitude ($Q/I{\approx}1.5\%$ as seen in Fig. 7) is an order of magnitude {\it larger}
than the observed one. This ``discrepancy'' is due to two reasons: (1) because
the result of Fig. 7 corresponds to a very simple model atom with only
three levels and (2) because the magnetic field has been assumed to be zero.
As shown in this volume
by Manso Sainz and Trujillo Bueno (2001a) the line-core amplitude
for the zero magnetic field reference case is reduced by about a factor 4
when a realistic five-level atomic model for Ca {\sc ii} is considered
(which includes the collisional coupling between the two {\it metastable} 
levels $^2{\rm D}_{3/2}$ and $^2{\rm D}_{5/2}$).
The remaining ``discrepancy''
is likely due to the Hanle effect of the chromospheric magnetic field, 
as illustrated in Section 7.  

Another interesting example of a group of lines having the same
angular momentum values
and showing a very similar observed
behaviour is given by the three lines of
multiplet no. 2 of Ba {\sc ii} (see the atlas of Gandorfer 2000;
and note that $82\%$ of the total barium abundance has no hyperfine structure splitting).

\subsection{Three lines with a common upper level: the Mg I {\boldmath $b$}
lines}

Consider the case of multiplet $^3{\rm P}^0\,-\,^3{\rm S}$, 
which leads to three spectral lines with a common upper level.
An interesting example is given by the three lines of
multiplet no. 2 of Mg I (i.e. the well-known Mg $b$ lines).
The Mg $b_4$ line at 5167 \AA$\,$ is a $J_l=0\,\rightarrow\,J_u=1$ transition
(with ${\cal W}=1$ and ${\cal Z}=0$; see Eq. 30), 
the Mg $b_2$ line at 5173 \AA$\,$ is a $J_l=1\,\rightarrow\,J_u=1$ transition
(with ${\cal W}={\cal Z}=-0.5$),
and the Mg $b_1$ line at 5184 \AA$\,$ is a $J_l=2\,\rightarrow\,J_u=1$ transition
(with ${\cal W}=0.1$ and ${\cal Z}{\approx}0.6$).

Stenflo {\it et al.} (2000) considered the Mg $b$ lines
``enigmatic'' because the observed linear polarization amplitudes
are {\it similar} (see also Stenflo {\it et al.} 1983), 
which is in total disagreement with the relative polarization amplitude scaling
they were expecting from the conventional polarization transfer theory which {\it neglects}
lower-level atomic polarization. 

Which is the {\it true} prediction of the conventional polarization transfer theory
that neglects lower-level atomic polarization?
Let us assume that the {\it exact} fractional atomic alignment of the common upper level
at line-core optical depth unity along the line of sight is $\sigma^2_0({u})=X>0$ so
as to yield, from Eq. (30), a {\it positive} linear polarization amplitude
for the Mg $b_4$ line (i.e. $(Q/I)_{b4} {\approx} X$), in agreement 
with the observations (recall that the Mg $b_4$ line has $J_l=0$ and
that $90\%$ of Mg has zero nuclear spin). If we now apply Eq. (30), but {\it neglecting}
the dichroism contribution coming from the atomic polarization of the
lower levels of the Mg $b_2$ and $b_1$ lines, we find $(Q/I)_{b2}
{\approx} -X/2$
(i.e. a {\it negative} polarization!) and $(Q/I)_{b1} {\approx} X/10$
(i.e. {\it ten times smaller} than that of the $b4$ line). However, as mentioned above,
the observations show {\it positive} polarization for the three lines,
and with similar amplitudes. 

\begin{figure}
\label{hanle9}
\plotone{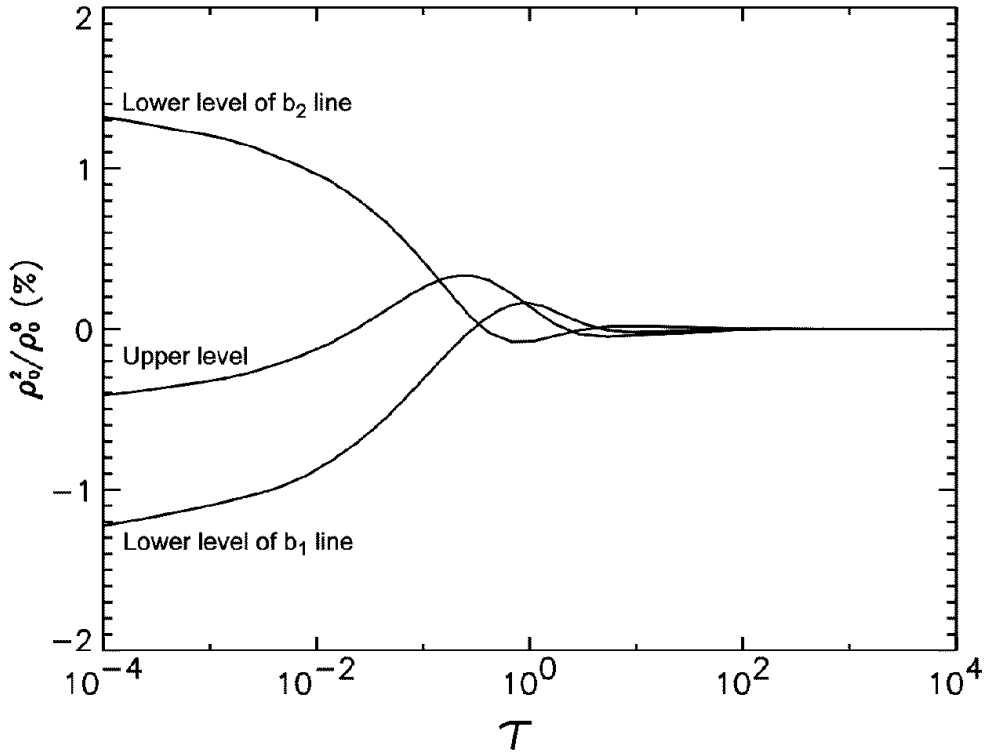}
\caption{The variation with the corresponding line optical depth 
of the fractional atomic polarization of the lower level of the
Mg $b_1$ line, of the lower level of the 
Mg $b_2$ line and of the upper level 
of the Mg $b_4$ line. Each of the three optical depths $\tau$ is
measured along the radial direction. The solar chromospheric model used is the VAL-C
model of Vernazza {\it et al.} (1981).}
\end{figure}

What could be the physical origin and explanation of the above-mentioned
observations? In a previous keynote article 
(cf. Trujillo Bueno 1999) I suggested that one could explain such 
spectropolarimetric observations
if we had (at line-core optical depth unity along the line of sight)
the following amounts of atomic polarization in the
metastable {\it lower levels} of the Mg $b_1$ and $b_2$ lines:

\begin{itemize}

\item For the lower level of the $b_2$ line $\,\rightarrow\,\,$ $\sigma^2_0({l})
\,{\approx}\,3X$.

\item For the lower level of the $b_1$ line $\,\rightarrow\,\,$ $\sigma^2_0({l})\,
{\approx}\,-2X$.

\end{itemize}
This ``explanation'' is based on the direct mapping that Eq. (30) implies:
the observed polarization amplitudes are giving us directly the atomic polarization
of the lower and upper levels at optical depth unity along the line of sight !

The crucial question now is: are these the atomic polarization values obtained
from self-consistent {\it multilevel} scattering polarization calculations 
in semi-empirical models of the solar chromosphere? The answer is affirmative,
as indeed results from a realistic multilevel coherency transfer calculation 
for a 19-level model atom for Mg I - Mg II (see Fig. 8). The results of this figure show
that the {\it metastable} lower-levels of the Mg I $b_1$ and $b_2$ lines are
{\it polarized} by the anisotropic radiation field of typical semi-empirical
solar chromospheric models.
The relative amplitudes of the calculated emergent polarizations 
of the three Mg $b$-lines at $\mu=0.1$ agree
fairly well with the observations, with {\it dichroism} playing again a critical role.

\section{The lower-level Hanle effect}

The lower-level $^2{\rm D}_{3/2}$ of the Ca {\sc ii} 8662 \AA $\,$ line
is {\it metastable}. Therefore, from Eq. (3), its atomic polarization and
the emergent linear polarization have to be sensitive to
milligauss fields. This is illustrated in Fig. 9, which
shows the self-consistent values of the $\rho^K_Q$-elements
for a horizontal field of 10 milligauss\footnote{These density-matrix 
values were calculated numerically
for the same three-level model atom mentioned in Section 6.1.}.
The assumed {\it horizontal}
field lies in the $xy$-plane of the reference system indicated in the
``north solar pole'' example of Figure 1.
It is orientated at 45 degrees (measured from $x$ towards $y$).
Fig. 9 shows the fractional population imbalances 
($\rho^2_0/\rho^0_0$) and the coherences ($\rho^2_Q/\rho^0_0$, with $Q=1$ and $Q=2$)
in percentage. Note that the 10 milligauss of the assumed horizontal field are sufficient
to reduce the population imbalances of the $^2{\rm D}_{3/2}$ level by 
a factor 3 approximately (compare with the thin solid line of Fig. 7). Moreover,
we now have non-zero coherences, {\it which are of the same order of magnitude
as the population imbalances themselves}. The emergent fractional linear
polarization ($Q/I$ and $U/I$) close to the stellar limb
($\mu=0.1$) and along the $x$-axis of Fig. 1 
(i.e. a direction of observation with azimuth $\chi=0$ and $\mu=0.1$)
is shown in Figure 10. Note that the $Q/I$ amplitude is a factor
3 smaller than the one corresponding to the zero magnetic field reference case of Fig. 7,
and that $U/I$ is of the same order of magnitude as $Q/I$.

\begin{figure}
\plotfiddle{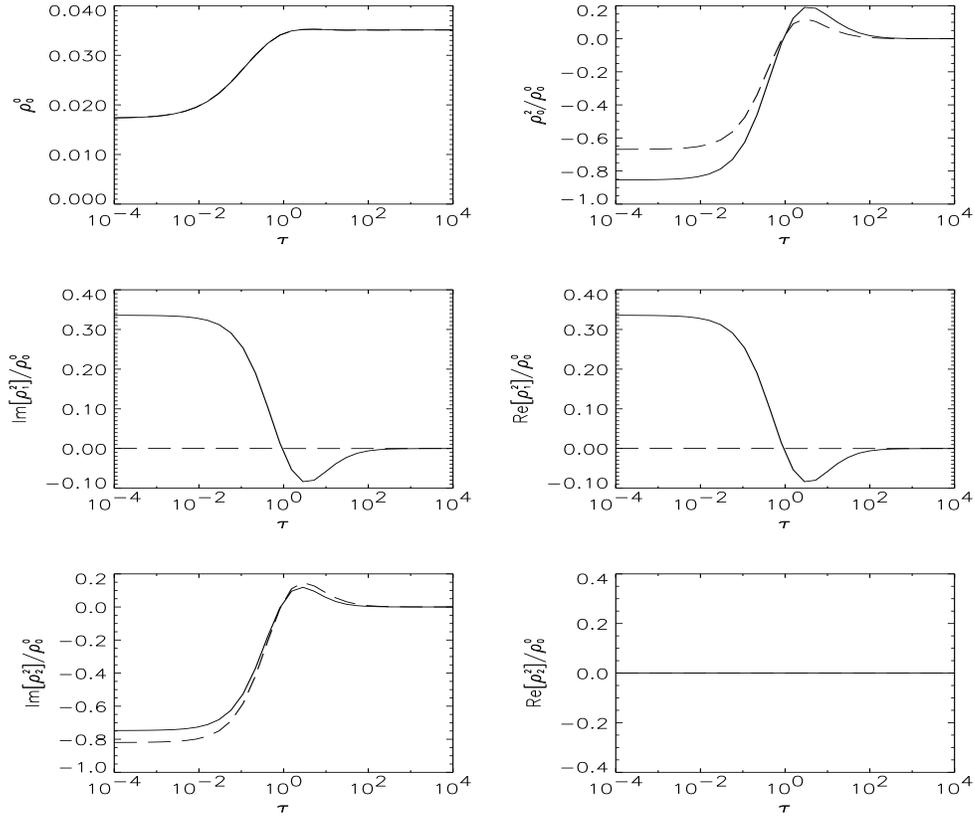}{10cm}{0}{75}{54}{-190}{-15}
\caption{\label{jbar}The $\rho^K_Q$-elements for a horizontal field of 10 milligauss
(solid lines) and of 10 gauss (dashed lines). 
Note that ${\rm Re}[\rho^2_2]/{\rho^0_0}=0$.}
\end{figure}
\begin{figure}
\plotfiddle{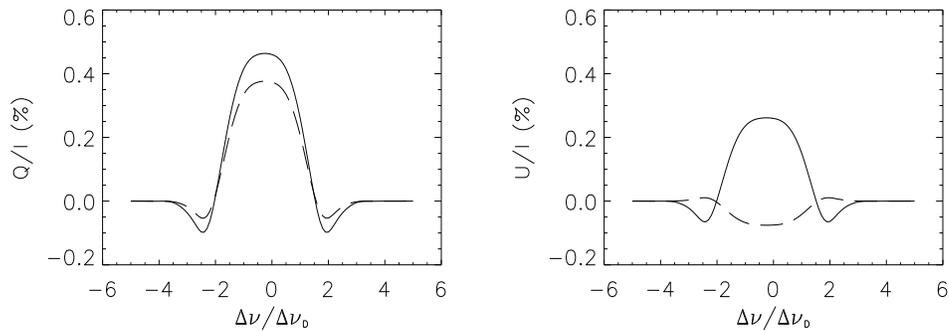}{6cm}{0}{75}{80}{-230}{-415}
\caption{The emergent $Q/I$ and $U/I$ of the 8662 \AA $\,$ line corresponding
to the two cases of Fig. 9. The simulated observation is
along the $x$-axis of Fig. 1 (i.e. at $\chi=0$, but at $\mu=0.1$) and for
an orientation of the assumed horizontal magnetic field as indicated in the text.}
\end{figure}

What is the physical origin of this $U/I$ profile?
It is exclusively due to the {\it coherences} of the
metastable {\it lower-level} of the Ca {\sc ii} 8662 \AA $\,$ line,
because $\epsilon_U=0$, while $\eta_U{\ne}0$ (see the paragraph after Eq. 25).
In fact, it is possible to derive Eddington--Barbier relations
for the $Q/I$ and $U/I$ emergent polarizations
due to the Hanle effect. This can be done
along similar lines to those which lead to Eq. (30) for the unmagnetized
reference case. Particularizing the Eddington--Barbier relation for $U/I$
to the direction of observation chosen to obtain the exact results 
of Fig. 10 (i.e. $\chi=0$) one obtains:

\begin{equation}
U/I\,{\approx}\,{\cal Z}\,\sqrt{3}\bigl{[}\sqrt{1-\mu^2}{\rm Im}(\sigma^2_1)+
\mu{\rm Im}(\sigma^2_2)\bigr{]},
\end{equation}
where $\sigma^2_1=\rho^2_1(l)/\rho^0_0(l)$ and $\sigma^2_2=\rho^2_2(l)/\rho^0_0(l)$ 
are the fractional coherences of the
{\it lower} level of the 8662 \AA $\,$ line at the spatial point
along the line of sight where the optical depth 
at the selected frequency is unity. Note that ${\cal Z}{\approx}0.7$, with 
${\cal Z}$ the symbol introduced after Eq. (30). Therefore, a positive
detection of a $U/I$ signal close to the limb ($\mu=0.1$) would be
a clean observational demonstration of the existence of
metastable-level {\it coherences}.   

Let us now consider what happens when
the intensity of the same horizontal field is 10 gauss.
As indicated by the dashed-lines of Figs. 9 and 10,
the $\rho^K_Q$-elements and the emergent linear polarization
are modified. The population imbalances of the $^2{\rm D}_{3/2}$ level
are not destroyed and some of its coherences remain.\footnote{For the case of
a microturbulent and isotropic magnetic field 
or for a microstructured horizontal field
with random azimuth, only the population imbalances
would remain (see the dashed-dotted line of Fig. 3 and the 
Appendix of the paper by Trujillo Bueno \& Manso Sainz 1999).}.
Note also that the emergent $Q/I$ in this saturated gauss regime
still has a sizeable amplitude ($Q/I {\approx} 0.4\%$), which is due to the
atomic polarization of the metastable $^2{\rm D}_{3/2}$ 
lower-level\footnote{We still find $Q/I\,{\approx}\,0.1\%$ 
when we use, instead of a three-level model, 
a realistic 5-level atomic model including the collisional coupling between the
metastable levels $^2{\rm D}_{3/2}$ and $^2{\rm D}_{5/2}$.}.
In other words, the atomic polarization of
long-lived atomic levels in the solar chromosphere
may still be sufficiently significant
(even after the partial destruction caused by a 10 gauss purely horizontal 
field with a deterministic or random azimuth!)
so as to be able to lead to emergent $Q/I$ signals in the observable range.
In a forthcoming publication we will show in detail that
a consistent explanation of the observed $Q/I$ in the {\it three} lines
of the Ca {\sc ii} IR triplet does {\it not} necessarily requires
to constrain the magnetic field to be either extremely low (with $B {\le} 10$ milligauss)
or vertically orientated (but with an intensity in the gauss regime). Note that I am
{\it not} saying that the chromospheric magnetic field is predominantly
horizontal. I am basically pointing out that some of the ``enigmatic'' features
of the {\it second solar spectrum} are due to metastable-level atomic
polarization and that the amount of this atomic polarization 
in topologically complex magnetic field scenarios with intensities in the gauss regime
can still be sufficiently significant so as to lead to the observed
$Q/I$ polarization amplitudes.
 
Finally, it may be found useful to distinguish between five
Hanle effect regimes in relation with the polarization of a spectral line
whose lower level is the ground or a metastable level. These Hanle regimes can
be established according to the values of $\Gamma_l$ and $\Gamma_u$ and
by noting that, for a given magnetic field intensity, $\Gamma_l\gg \Gamma_u$
(because the lower-level lifetime is about two orders of magnitude
larger than the upper-level lifetime). As mentioned earlier,
in this section we are using the reference system
whose $z$-axis is directed along the radial direction 
of the star:\footnote{See Landi Degl'Innocenti (1985) and Section 5.2 for a discussion
of Hanle-effect regimes from the point of view of the magnetic field reference frame.}

\begin{itemize}

\item {\bf Regime 1:} $\Gamma_l\,\ll \,1\,\,\,$ and $\,\,\,\Gamma_u\,\ll\, 1$

In this regime the atomic alignment
of the lower level of the line transition of interest is typically of the
same order of magnitude as the upper-level alignment (i.e.
$|\sigma^2_0(l)| \sim |\sigma^2_0(u)|$). We find here the maximum
possible $Q/I$ amplitude.

\item {\bf Regime 2:} $\Gamma_l\,{\approx}\,1\,\,\,$ and $\,\,\,\Gamma_u\,\ll \, 1$

In this regime the emergent
polarization is sensitive to milligauss fields.

\item {\bf Regime 3:} $\Gamma_l\,\gg \,1\,\,\,$ and $\,\,\,\Gamma_u\,\ll \, 1$

In this regime we have $|\sigma^2_0(l)| \ll |\sigma^2_0(u)|$.

\item {\bf Regime 4:} $\Gamma_l\,\gg \,1\,\,\,$ and $\,\,\,\Gamma_u\,{\sim}\, 1$

In this regime the emergent
polarization is sensitive to magnetic fields in the gauss regime.

\item {\bf Regime 5:} $\Gamma_l\,\gg \,1\,\,\,$ and $\,\,\,\Gamma_u\,
\gg 1\,$

We again find that $|\sigma^2_0(l)| \sim |\sigma^2_0(u)|$,
but with lower individual values for $|\sigma^2_0(l)|$ and $|\sigma^2_0(u)|$
than for the unmagnetized reference case. 
The emergent polarization is sensitive only to the field orientation.
In this regime, one should find
relative amplitudes of the linear polarizations observed
in multiplets (e.g. in the Mg I $b$ lines) similar to those found
for the unmagnetized reference case. This emphasizes the importance
of an accurate quantification of the observed polarization amplitudes
in order to infer correctly the intensity and inclination of the magnetic field. 

\end{itemize}

\section{Concluding remarks}

This review article, besides providing an introduction
to optical pumping, atomic polarization and the Hanle effect,
has advanced some new results which we will publish in
suitable journals during the following months. 
Of particular interest for the reader who looks at the Sun as a unique
physics laboratory is the conclusion that
{\it the quiet solar chromosphere is a polarized vapor
with optical properties similar to those of an anisotropic crystal}.

On the other hand, the reader interested mainly 
in the remote sensing of solar and stellar
magnetic fields should feel optimistic because
the quest for understanding the physical origin of the ``second solar
spectrum'' is now becoming  one of the success stories of astrophysics.
We had the theory (cf. Landi Degl'Innocenti 1983). We had the
observations (cf. Stenflo and Keller 1997; Stenflo {\it et al.} 2000).
And now we have the {\it self-consistent} radiative transfer
modeling, which is the essential link between
theory and observations. In the following years rigorous confrontations
of observations of weak polarization signals with numerical simulations
of scattering polarization and of the Hanle and Zeeman effects should lead to
fundamental new insights in our understanding of solar photospheric and
chromospheric magnetism. 

\acknowledgments
The author wishes to express his gratitude 
to Egidio Landi Degl'Innocenti and Rafael Manso Sainz for their 
collaboration and for valuable inputs and discussions, and
to Michael Sigwarth for his invitation to participate in a highly interesting
workshop. This work is part of the EC-TMR European Solar Magnetometry Network.

\end{document}